\documentclass[a4paper,fleqn,usenatbib]{mnras}

\usepackage{savesym}
\usepackage{amsmath}
\savesymbol{iint}
\usepackage{txfonts}
\restoresymbol{TXF}{iint}
\usepackage{lastpage}
\usepackage{mathtools}
\usepackage[T1]{fontenc}
\usepackage{ae,aecompl}
\usepackage{hyperref}
\usepackage{amssymb}
\usepackage{url}
\usepackage{gensymb}
\usepackage{subfigure}
\usepackage{wasysym}
\usepackage{textcomp}
\usepackage{multirow}
\usepackage{multicol}
\usepackage[version=3]{mhchem}
\usepackage{txfonts}
\usepackage{epstopdf}

\usepackage{graphicx}	
\usepackage{amsmath}	
\usepackage{amssymb}	
\usepackage{color}
\usepackage{comment}
\usepackage{url}

\title[Hot Jupiter composition]{The composition of hot Jupiter atmospheres assembled within chemically evolved protoplanetary discs}

\author[S. Notsu et al.]{
Shota Notsu$^{1,2}$\thanks{E-mail: shota.notsu@riken.jp}\thanks{RIKEN Special Postdoctoral Researcher (SPDR, Fellow)},
Christian Eistrup$^{3,2}$\thanks{Virginia Initiative on Cosmic Origins (VICO) Postdoctoral Fellow}
\thanks{S.~Notsu and C.~Eistrup contributed equally to this paper.},
Catherine Walsh$^{4}$,
and
Hideko Nomura$^{5}$ 
\\
$^{1}$Star and Planet Formation Laboratory, RIKEN Cluster for Pioneering Research, 2-1 Hirosawa, Wako, Saitama 351-0198, Japan\\
$^{2}$Leiden Observatory, Faculty of Science, Leiden University, PO Box 9513, 2300 RA Leiden, The Netherlands\\
$^{3}$Department of Astronomy, University of Virginia, 530 McCormick Rd, Charlottesville, VA 22904, USA\\
$^{4}$School of Physics and Astronomy, University of Leeds, Leeds, LS2 9JT, UK\\
$^{5}$National Astronomical Observatory of Japan, 2-21-1 Osawa, Mitaka, Tokyo 181-8588, Japan
}

\date{Accepted 2020 September 18. Received 2020 August 31; in original form 2020 June 22}

\pubyear{2020}

\begin{document}
\label{firstpage}
\pagerange{\pageref{firstpage}--\pageref{LastPage}}
\maketitle

\begin{abstract}
The radial-dependent positions of snowlines of abundant oxygen- and carbon-bearing molecules in protoplanetary discs will result in systematic radial variations in the C/O ratios in the gas and ice. 
This variation is proposed as a tracer of the formation location of gas-giant planets.
However, disc chemistry can affect the C/O ratios in the gas and ice, thus potentially erasing the chemical fingerprint of 
snowlines in gas-giant atmospheres.
We calculate the molecular composition of hot Jupiter atmospheres
using elemental abundances extracted from a chemical kinetics model of a disc midplane where we have varied the initial abundances and ionization rates.
The models predict a wider diversity of possible atmospheres than those predicted using elemental ratios from snowlines only.
As found in previous work, as the C/O ratio exceeds the solar value, the mixing ratio of \ce{CH4} increases in the lower atmosphere, and those of \ce{C2H2} and \ce{HCN} increase mainly in the upper atmosphere. 
The mixing ratio of \ce{H2O} correspondingly decreases. 
We find that hot Jupiters with C/O$>1$ can only form between the \ce{CO2} and \ce{CH4} snowlines.
Moreover, they can only form in a disc which has fully inherited interstellar abundances, and where negligible chemistry has occurred.
Hence, carbon-rich planets are likely rare,
unless efficient transport of hydrocarbon-rich ices via pebble drift 
to within the \ce{CH4} snowline is a common phenomenon.
We predict combinations of C/O ratios and elemental abundances that can constrain gas-giant planet formation locations relative to snowline positions, and that can provide insight into the disc chemical history. 
\end{abstract}

\begin{keywords}
astrochemistry--- protoplanetary discs--- ISM: molecules--- planets and satellites: gaseous planets--- planets and satellites: atmospheres
\end{keywords}


\section{Introduction}
Exoplanets are ubiquitous. 
Now, twenty-five years after the discovery of the first exoplanet around a main-sequence star \citep{Mayor1995}, 
it is estimated that for every star in the galaxy, there is at least one exoplanet \citep{Winn2015}.
From detecting these exoplanets, and determining their sizes, masses and orbital characteristics, exoplanetary science has now moved to measuring the atmospheric compositions of exoplanets. 
Detections of simple molecules (such as CO, \ce{H2O}, \ce{CH4}, and \ce{HCN}) have been reported in recent years, mostly in the atmospheres of giant hot Jupiters (e.g., \citealt{Swain2008, Snellen2010, Stevenson2010, Madhusudhan2011, Waldmann2012, Konopacky2013, Moses2013, Kreidberg2014, Brogi2016, Birkby2017, MacDonald2017, Samland2017, Hawker2018, Cabot2019, Guilluy2019, Madhusudhan2019}). 
However, for many hot Jupiters, the error bars on derived abundances are too large to constrain characteristics related to their formation \citep{Brewer2017}.
Improved constraints on the abundances in exoplanet atmospheres require the next generation of facilities, such as JWST, ARIEL, SPICA, and future ground-based telescopes (e.g., E-ELT and TMT). These observatories will measure exoplanet atmospheric gas abundances to levels that may enable chemical differentiation between exoplanets, including constraining their atmospheric carbon-to-oxygen (C/O) ratios with a sufficiently high precision to discriminate between formation mechanisms and formation locations within the protoplanetary disc (e.g., \citealt{Kawahara2014, Greene2016, Greene2019, Schlawin2018, Tinetti2018, Venot2018, Bowler2019, Brogi2019, Madhusudhan2019, Changeat2020, Venot2020}).

In preparation for observations with these future facilities, it is necessary to improve our understanding of how different atmospheric compositions relate to the environments in which exoplanets form, namely the midplanes of protoplanetary discs. Here, it is important to understand the origin and variation in chemical composition of the planet-forming material in disc midplanes, and to quantify how disc midplane chemistry during the epoch of planet formation influences exoplanetary compositions.

Exoplanetary cores are thought to be built from the solid material (refractory rocks, coated in volatile ices if beyond the water snowline) in the disc midplane, while exoplanet atmospheres are built from accreted gas, and from the volatile contents of solids from impacting planetesimals (e.g., \citealt{Madhusudhan2014, Madhusudhan2019}). 
Thus, the chemical composition of the gas and ice in the disc midplane is expected to influence the composition of exoplanet atmospheres. Of particular focus has been the variation in elemental abundances in gas and ice due to the locations of major snowlines of carbon-, oxygen-, and nitrogen-bearing molecules \citep{Oberg2011}. 
Several recent works have attempted to predict the exoplanet atmospheric compositions from assumptions about the chemical composition in the disc midplanes induced by physical mechanisms prior to the onset of planet formation (e.g., \citealt{Oberg2016, Piso2016, Booth2017, Booth2019, Madhusudhan2017}). 
These works have greatly improved our understanding of the physical effects leading to the formation of exoplanets and their atmospheres (due to effects such as grain growth, particle coagulation, migration and diffusion, vertical mixing, and the effects of volatile snowlines on grain evolution).
An important next step is to determine how the chemistry of the volatile material in the planet-building zone may change its composition and thus the composition of the atmospheres of nascent forming planets \citep{Booth2019, Cridland2019a}.

Many studies, including \citet{Aikawa1997}, \citet{ChaparroMolano2012}, \citet{Helling2014}, \citet{Walsh2015}, \citet{Eistrup2016, Eistrup2018}, \citet{Notsu2016, Notsu2017}, \citet{Cridland2017, Cridland2019a, Cridland2019b, Cridland2020}, \citet{Yu2017}, \citet{Bosman2018} and \citet{Booth2019} have modelled chemical evolution in disc midplanes prior to planet formation.
\citet{Cridland2017, Cridland2019a, Cridland2019b, Cridland2020} and \citet{Booth2019} used the results of such chemical evolution models as input for their planet formation and exoplanet atmosphere model. 
However, the types and numbers of chemical reactions included in the chemical models in these works vary across the models; hence, the dependence of the results on the chemical parameters and setups are not straightforward to evaluate. 
The ice chemistry, taking place on grain surfaces, is particularly challenging to model since the reactions between ice species are less constrained by laboratory experiments than gas-phase reactions \citep[see, e.g.,][]{Cuppen2017}. Also various assumptions need to be made about the sizes and shapes of dust grains, and the ability of atoms and molecules to move around on and within the ice. 
In \citet{Eistrup2016} disc midplane chemistry was modeled with and without detailed ice chemistry, and indeed important differences were seen when including the ice chemistry, with the formation of water ice particularly affected.
Further also explored in that work was the impact of considering different starting initial abundances and different levels of ionisation.
Since that work, there have been several other studies that include (some) ice chemistry in disc midplane models \citep[e.g., ][]{Cridland2017, Bosman2018, Cridland2019a, Cridland2019b, Booth2019, Cridland2020}.
Of these, the models of \citet{Bosman2018}, \citet{Cridland2019a}, and \citet{Cridland2020} adopt a chemistry of a similar complexity as in \citet{Eistrup2016}.

The investigation undertaken in \citet{Eistrup2016} led to estimations of the C/O ratios in gas and ice in the disc midplane as a function of radius in the disc. 
These C/O ratio profiles (see Figure 6 of \citealt{Eistrup2016}) varied depending on the model setup, and generally showed different trends from those in the ``stepfunction''-picture of the C/O ratios presented in \citet{Oberg2011}. In the work of \citet{Oberg2011}, the C/O ratios in the gas and for the ices on the grains were estimated using the snowlines of the main volatiles, thus considering freeze-out and desorption only. 

The goal of this paper is to 
compute the atmospheric composition of hot Jupiter atmospheres using the C/O ratios and elemental abundances of gas-phase volatiles extracted from chemical kinetics models of protoplanetary disc midplanes \citep{Eistrup2016}.  
We assume that the hot Jupiter has formed via the core accretion mechanism and that it has accreted its atmosphere locally and only from the gas in the vicinity of the planetary core \citep{Pollack1996, Ikoma2000}. 
Hence, we consider that such planets are located within a narrow radial region during the runaway acquisition of their atmospheres (see, e.g., \citealt{Mordasini2012}, \citealt{Alessi2017}, and \citealt{Cridland2019b}, for more details).
We further assume that the gas giant has migrated inwards to its current location post its formation in the outer disc (e.g., \citealt{Lin1986, Ida2004, Ida2018, Hasegawa2012, Kanagawa2018}, see also Section~4.4).  
Thus the composition of the atmosphere of the gas giant planet is set by the gas-phase composition in the disc midplane at the location of its formation.  
In this study we will quantify, for the first time, the composition of hot Jupiter atmospheres that have been accreted from a protoplanetary disc in which significant chemical evolution has taken place. We will compare the atmospheric mixing ratios of key volatiles (such as CO, \ce{H2O}, and \ce{CH4}) computed using chemically evolved disc midplane material with those predicted by simple models of disc midplane composition (i.e., those without chemistry) to assess the effect of disc midplane chemical evolution on exoplanet atmospheric compositions.

\section{Methods}
\subsection{Physical structure of hot Jupiter atmospheres}
The dayside atmospheric structure of a hot Jupiter used in this paper is calculated using the methods described in \citet{Guillot2010}.
\citet{Guillot2010} presents an analytical description of the 1D pressure-temperature structure of an irradiated planetary atmosphere
under the assumptions that radiative equilibrium holds and that the atmosphere can be described as a plane-parallel gray atmosphere.
The atmospheric temperature structure is determined by the following equations 
(see also equation 27 and Section 2 of \citealt{Guillot2010}),

\begin{align*}
T^{4} & = \frac{3T^{4}_{\mathrm{int}}}{4}\left(\frac{2}{3} + \tau\right)  \\          
& +\frac{3T^4_{\mathrm{irr}}}{4}\mu_{*}\left[\frac{2}{3}\text{+}\frac{\mu_{*}}{\gamma}\text{+}\left(\frac{\gamma}{3\mu_{*}}-\frac{\mu_{*}}{\gamma}\right)\exp\left\{-\frac{\gamma\tau}{\mu_{*}}\right\}\right],\,\, \text{where}\\
T_{\mathrm{irr}}
& = T_{\mathrm{\star}}\left(\frac{R_{\mathrm{\star}}}{D}\right)^{0.5}, \\
\tau & =  m\,\kappa_{\mathrm{ir}}, \\
\gamma & = \kappa_{\mathrm{vis}}/\kappa_{\mathrm{ir}}, \,\, \text{and}\\
\mu_{*}& = \cos\theta_{*}.
\end{align*}

Here, we assume the solar effective temperature, $T_{\mathrm{\star}}=5778$~K, and radius, $R_{\mathrm{\star}}=\mathrm{R}_{\odot}$.
We assume that the planet receives a flux equal to $\sigma T^{4}_{\mathrm{irr}}$ from its parent star, and that it emits an intrinsic heat flux $\sigma T^{4}_{\mathrm{int}}$ (we adopted $T_{\mathrm{int}}=100$~K).

The semi-major axis of the planetary orbit is $D$, and we consider here two cases: a hot Jupiter that has migrated to 0.05~au, and one that has migrated to 0.1~au.
According to previous observational results (e.g., \citealt{Wright2011, Winn2015}), the semi-major axes of hot Jupiters are typically $\sim 0.01 - 0.1$ au.
The optical depth of the atmosphere is $\tau$, and $\theta_{*}$ corresponds to the angle between the direction of incidence 
of stellar irradiation and the local vertical: we adopt $\theta_{*} = 0$.
The column mass from the top of the atmosphere downwards is given by $m$ where $dm = \rho\,dz$, where $\rho$ is the mass density and $z$ is the depth into the atmosphere. 
The mass absorption coefficients, $\kappa_{\mathrm{vis}}$ and $\kappa_{\mathrm{ir}}$ are those at visible and infrared wavelengths, respectively. 
We assume that absorption of stellar radiation occurs mainly at visible wavelengths, and that (re)emission occurs at infrared wavelengths only.  
We adopt $\kappa_{\mathrm{vis}} =4.0 \times10^{-3}$~cm$^{2}$~g$^{-1}$ and $\kappa_{\mathrm{ir}} = 1.0 \times10^{-2}$~cm$^{2}$~g$^{-1}$, which well reproduce 
the detailed temperature-pressure profiles of well-studied hot Jupiters such as HD~209458~b \citep{Guillot2010}.

\begin{figure}
\hspace{0.0cm}
\includegraphics[width=0.5\textwidth]{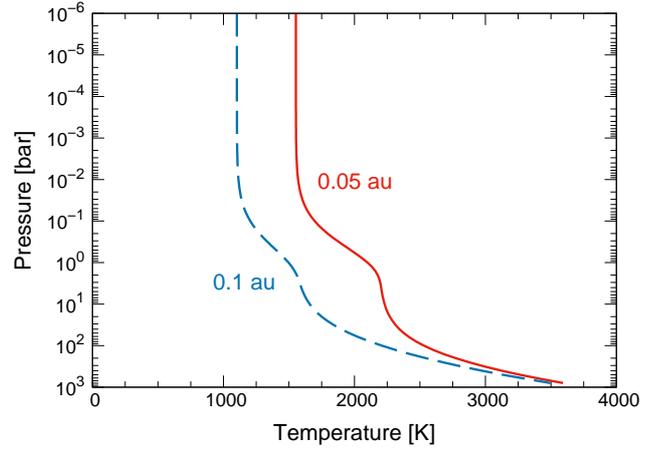}
\vspace{0.3cm}
\caption{The temperature-pressure profiles of two hot Jupiter atmospheres at an orbital distance of 0.05 au (red solid curve) and 0.1 au (blue dashed curve). 
Both profiles are calculated using the analytical formula presented in \citet{Guillot2010}.}
\label{Figure1}
\end{figure}

Figure \ref{Figure1} shows the calculated temperature and pressure profiles for the two cases studied here where the present planet position is 0.05 au (red curve) and 0.1 au (blue curve).
The temperature of both atmospheres exceeds 1000~K at all heights. In addition, the temperature at 1 bar is $\approx~2100$~K and 1500~K, respectively.

\subsection{Calculating the chemical structure of hot Jupiter atmospheres}
We calculate the chemical structure of hot Jupiter atmospheres assuming that chemical equilibrium holds in the region probed by near- to mid-infrared spectroscopy.
In our adopted physical structure, the value of atmospheric pressure at $\tau\sim 1$ in the infrared wavelength range is around 0.3~bar; thus, infrared observations mainly trace the region with $\sim 0.1-1$~bar.
According to previous studies (e.g., \citealt{Moses2011, Line2013}), in hot exoplanetary atmospheres ($T > 1200$~K), thermochemical equilibrium dominates over disequilibrium effects, such as photochemistry. 
In addition, for the hottest planets ($T \gtrsim 1200$~K at 0.1 bar), major molecules (such as \ce{H2O}, \ce{CO}, \ce{CH4}, and \ce{H2}) are predicted to be in thermochemical equilibrium even under a wide range of vertical mixing strengths \citep{Line2013, Line2014}.

The open-source Thermochemical Equilibrium Abundances (TEA\footnote[1]{https://github.com/dzesmin/TEA}; \citealt{Blecic2016}) code is used to compute the mixing ratios in the atmosphere using the temperature-pressure profiles shown in Figure~\ref{Figure1}.
The TEA code adopts the methodology of \citet{White1958} to perform the Gibbs free-energy minimization necessary for calculating the thermochemical equilibrium abundances.
Given a temperature and pressure and an elemental composition, the TEA code determines the set of mole fractions of the desired gaseous molecules that minimizes the total chemical potential of the system.
In our calculations, we include the following gaseous molecules; \ce{H}, \ce{H2}, \ce{He}, \ce{C}, \ce{O}, \ce{N}, \ce{CO}, \ce{CO2}, \ce{CH4}, \ce{H2O}, \ce{N2}, \ce{HCN}, \ce{NH3}, \ce{C2H2}, and \ce{C2H4}, many of which are expected to be detected in hot-Jupiter atmospheres with infrared spectroscopic observations.
Here we note that \citet{Heng2016} also developed an analytical method for computing the abundances of major molecules in a C-H-O-N system in chemical equilibrium, which reproduce well the results of the TEA code.

To create a set of elemental abundances for input to the TEA code, we extracted the elemental composition of the gas in a protoplanetary disc midplane in which chemical evolution has taken place over 1~Myr and in which we have explored the impact of the assumed initial abundances and cosmic-ray ionisation rate on the gas composition  \citep[see][for full details]{Eistrup2016}. 
We extract the elemental composition (C/H, O/H, N/H) at 0.5~au, 1.0~au, 5.0~au, and 20.0~au from the central star (see Table \ref{tab:1}).
These radii correspond to the following radial regions: i) inside the water snowline, ii) between the water and \ce{CO2} snowlines, iii) between the \ce{CO2} and \ce{CH4} snowlines, and iv) between the \ce{CH4} and \ce{CO} snowlines.
The snowline positions of \ce{H2O}, \ce{CO2}, \ce{CH4}, and CO in our adopted disc model \citep{Eistrup2016} are 0.7, 2.6, 16, and 26~au, at temperatures of 177, 88, 28, and 21~K, respectively.

We do not consider a case beyond the \ce{CO} snowline because here the gas is completely depleted in both C and O.
However, we note that \citet{Bosman2019} and \citet{Oberg2019} have both suggested that the core of Jupiter could have formed at or outside the \ce{N2} snowline ($> 30$~au) evidenced 
by uniform abundance patterns of several elements including nitrogen. 
\citet{Oberg2019} discussed that once the core has formed and migrated, all observed elemental abundances can be explained by Jupiter having accreted the bulk of its gaseous envelope at smaller radii. We note here that in this work we adopt the assumption that the bulk of the atmosphere is accreted from a narrow radial region within the disc during the runaway accretion phase.
%

\subsection{Elemental abundances and C/O ratios in chemically evolved protoplanetary discs}
\begin{figure*}
\includegraphics[width=0.45\textwidth]{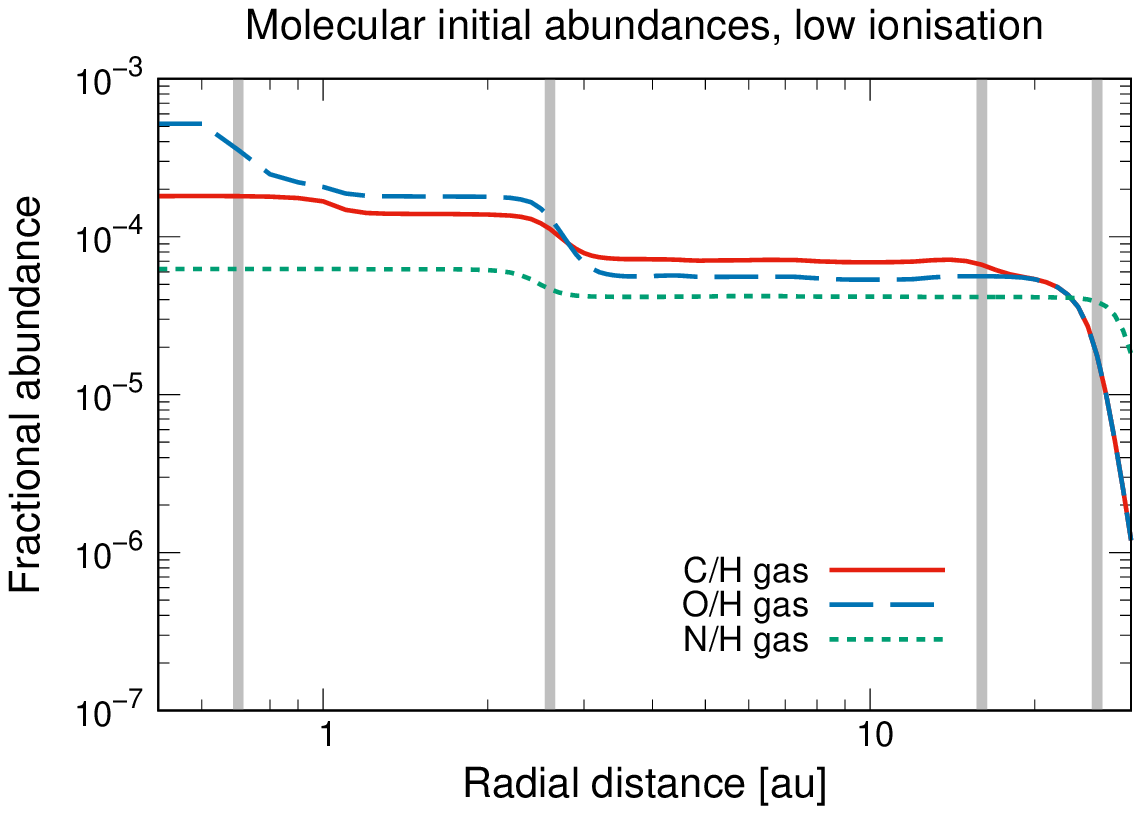}
\includegraphics[width=0.45\textwidth]{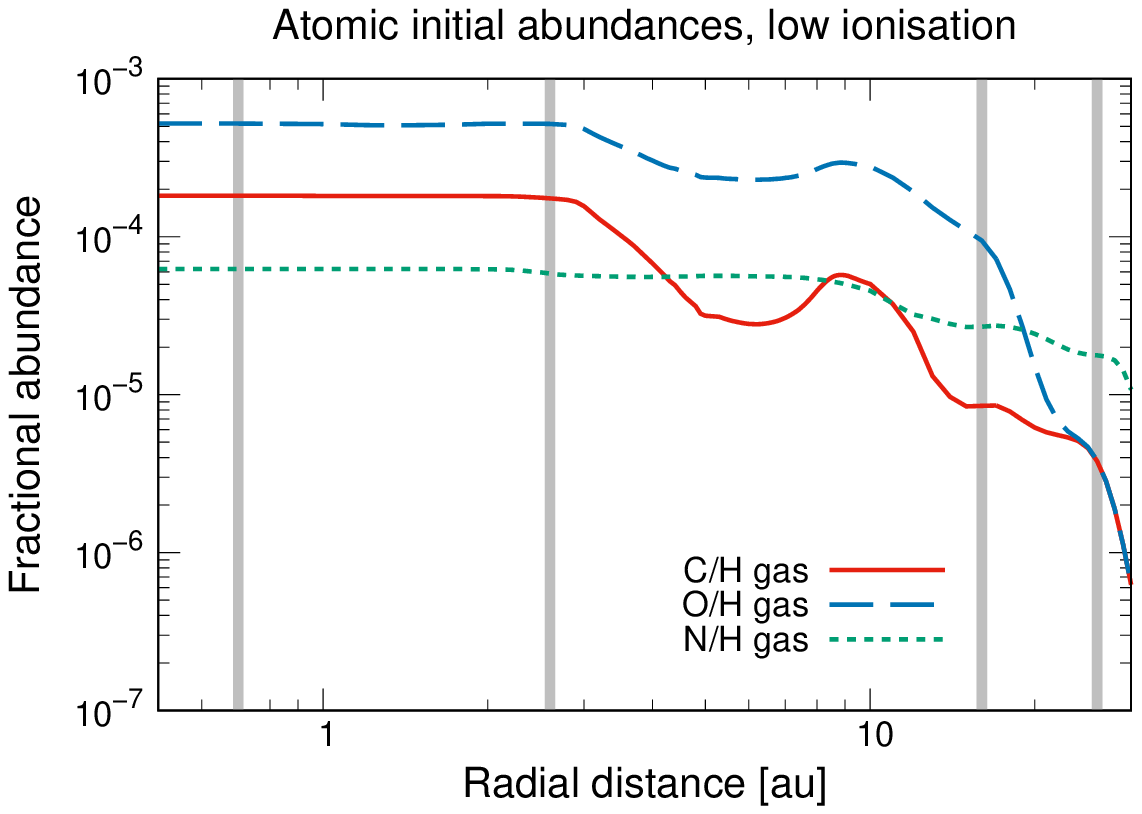}
\includegraphics[width=0.45\textwidth]{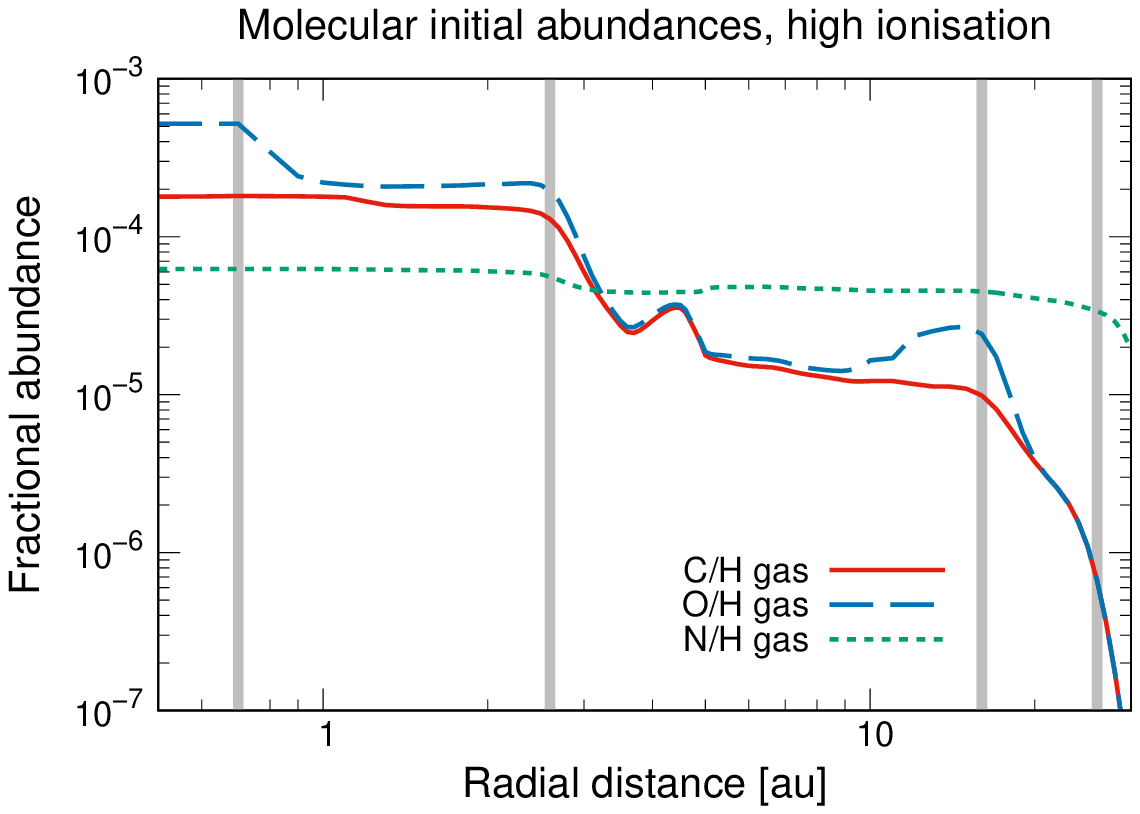}
\includegraphics[width=0.45\textwidth]{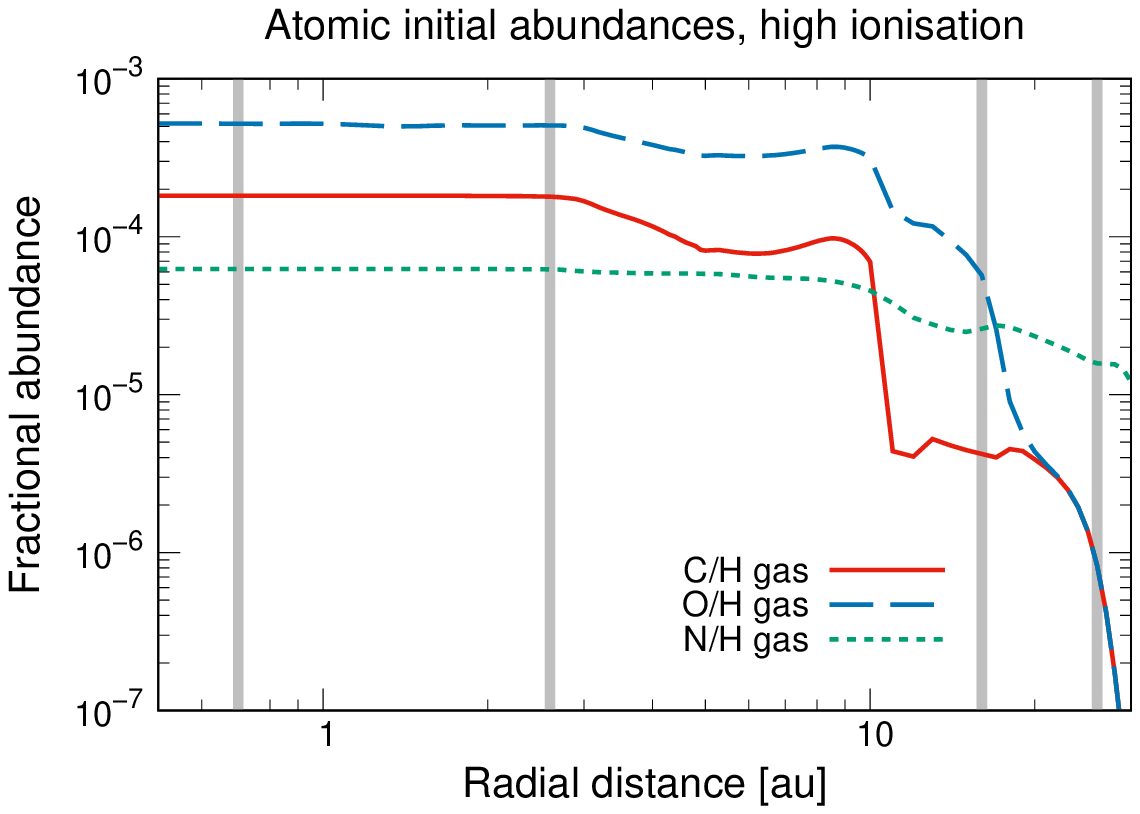}
\vspace{0.5cm}
\caption{Disc gas-phase elemental abundances with respect to total H nuclei density as a function of radial distance from the central star for key elements, carbon (red solid lines), oxygen (blue dashed lines), and nitrogen (green dotted lines). These were obtained from the results using the full chemical network in \citet{Eistrup2016}. 
The top two panels show the results for the low ionisation case ($\zeta<10^{-18}$ s$^{-1}$) and the bottom two panels show those for the high ionisation case ($\zeta\sim10^{-17}$ s$^{-1}$).
The left-hand panels show the results when assuming molecular initial abundances (the ``inheritance" scenario) and the right-hand panels show those when
assuming atomic initial abundances (the ``reset" scenario).
The gray vertical bands show the positions of snowlines for key volatiles (0.7, 2.6, 16 and 26 au, for \ce{H2O}, \ce{CO2}, \ce{CH4}, and CO, respectively).}
\label{Figure2_new2}
\end{figure*}

Figure \ref{Figure2_new2} shows
the gas-phase elemental abundance with respect to total H nuclei density in the 1 Myr disc midplane as a function of radial distance from the central star for carbon, oxygen, and nitrogen, 
were obtained from the abundances of key volatiles calculated using the full chemical network in \citet{Eistrup2016}. 
We reproduce those data here in a different form than published previously to enable a comparison between the elemental ratios predicted by the four considered disc models.
The grey vertical bands represent the positions of the snowlines of \ce{H2O}, \ce{CO2}, \ce{CH4}, and \ce{CO}. 
Presented are the calculated gas-phase elemental abundances for four different models in which either atomic or molecular initial abundances were assumed, and in which either a high ($\zeta\sim10^{-17}$ s$^{-1}$) or low ($\zeta<10^{-18}$ s$^{-1}$) level of cosmic-ray ionisation was assumed. 
The latter case assumes that the young (T Tauri) star's magnetic field can deflect galactic cosmic rays \citep[see, e.g.,][]{Cleeves2013}. 
The use of atomic initial abundances implies that full chemical reset has occurred during disc formation via, e.g., shocks generated during accretion from the protostellar envelope onto the disc.  
On the other hand, the use of molecular initial abundances assumes that the disc material is wholly inherited from the molecular cloud from which the central star formed, implying a more quiescent mode of disc formation.
The atomic and molecular initial abundances at $t = 0$, and representing these ``reset" and ``inheritance" scenarios respectively, are listed in table~1 in \citet{Eistrup2016}.
%

Table \ref{tab:1} displays the adopted gas-phase elemental abundances that were used in the calculation of the composition of hot Jupiter atmospheres and which were extracted from the data presented in Figure \ref{Figure2_new2}.
We provide these in tabular form to allow these data to be used in future works, and to facilitate reproduction of the atmospheric simulations by other researchers.
The inclusion of chemistry has a significant impact on the C/O ratios in both gas and ice in the planet-forming region which is expected to influence the resulting composition of planetary atmospheres.
In the model with molecular initial abundances (the ``inheritance" scenario) and a low ionisation, 
the chemical evolution effects are negligible, and the behaviour of the elemental abundances with radius are similar to those found in the ``stepfunction"  picture presented in \citet{Oberg2011}. 
The gas-phase C/O ratio increases as the \ce{H2O} and \ce{CO2} snowlines are surpassed and oxygen is removed from the gas, 
and it exceeds 1.0 between the \ce{CO2} and CO snowlines.
In the model with molecular initial abundances and a high ionisation, between $\approx 1$ and 15 au, gas-phase \ce{CH4} is efficiently converted to \ce{CO2} via chemistry induced by cosmic-ray induced photons. 
In addition, gas-phase \ce{O2} is also efficiently produced.
Thus the C/O ratio is $< 1$ between the \ce{CO2} and CO snowlines in this case.
The ices remain, on the whole, dominated by oxygen (i.e., C/O~$ < 1$) for the inheritance scenario. 

\begin{table*}
	\centering
	\caption{The elemental abundances used in the computation of the composition of hot Jupiter atmospheres (from \citealt{Eistrup2016}).}
	\label{tab:1}
	\begin{tabular}{lccccc} 
                \hline
		\multicolumn{6}{c}{Molecular initial abundances} \\
		\hline 

	        && C/H & O/H & N/H & C/O \\
		\hline
		&0.5 au & 1.81$\times10^{-4}$ & 5.20$\times10^{-4}$ & 6.24$\times10^{-5}$& 0.35\\
		low&1 au & 1.67$\times10^{-4}$ & 2.07$\times10^{-4}$ & 6.24$\times10^{-5}$& 0.81\\
		ionisation&5 au & 7.06$\times10^{-5}$ & 5.57$\times10^{-5}$ & 4.18$\times10^{-5}$& 1.27 \\
                &20 au& 5.40$\times10^{-5}$ & 5.37$\times10^{-5}$ & 4.14$\times10^{-5}$&1.00 \\
		&&  &  & &\\	
		
		&0.5 au & 1.80$\times10^{-4}$ & 5.19$\times10^{-4}$ & 6.24$\times10^{-5}$&0.35\\
		high&1 au & 1.80$\times10^{-4}$ & 2.20$\times10^{-4}$ & 6.24$\times10^{-5}$&0.82\\
		ionisation&5 au & 1.76$\times10^{-5}$ & 1.87$\times10^{-5}$ & 4.68$\times10^{-5}$&0.94\\
                &20 au& 3.07$\times10^{-6}$ & 3.07$\times10^{-6}$ & 3.99$\times10^{-5}$&1.00 \\
                 
		\hline	
		\multicolumn{6}{c}{Atomic initial abundances} \\
		\hline 

		&& C/H & O/H & N/H & C/O \\
		\hline
		&0.5 au & 1.81$\times10^{-4}$ & 5.21$\times10^{-4}$ & 6.24$\times10^{-5}$&0.35\\
		low&1 au & 1.81$\times10^{-4}$ & 5.17$\times10^{-4}$ & 6.24$\times10^{-5}$&0.35\\
		ionisation&5 au & 3.16$\times10^{-5}$ & 2.37$\times10^{-4}$ & 5.65$\times10^{-5}$&0.13\\
                &20 au& 6.19$\times10^{-6}$ & 1.51$\times10^{-5}$ & 2.43$\times10^{-5}$&0.41 \\
		&&  &  & &\\	
		
		&0.5 au & 1.81$\times10^{-4}$ & 5.21$\times10^{-4}$ & 6.24$\times10^{-5}$&0.35\\
		high&1 au & 1.81$\times10^{-4}$ & 5.19$\times10^{-4}$ & 6.24$\times10^{-5}$&0.35\\
		ionisation&5 au & 8.16$\times10^{-5}$ & 3.25$\times10^{-4}$ & 5.82$\times10^{-5}$&0.25\\
                &20 au& 3.90$\times10^{-6}$ & 4.36$\times10^{-5}$ & 2.34$\times10^{-5}$&0.89 \\
		\hline	
	\end{tabular}
\end{table*}

In the models with atomic initial abundances (the ``reset" scenario), 
CO, \ce{O2}, and atomic oxygen are the main gas-phase carriers of carbon and oxygen outside the water snowline.
Within 1 Myr, the formation of gas-phase \ce{O2} is faster than the formation of \ce{H2O} ice: 
\ce{O2} is very volatile and only freezes out at 24~K.
Thus, the gas-phase C/O ratios are $< 1.0$ at all radii considered here for the case of chemical reset.
In addition, the ice is more carbon-rich in this scenario than in the inheritance scenario, although still remaining oxygen rich (C/O~$<1$; \citealt{Eistrup2016}).

The data presented in Table~\ref{tab:1} demonstrate well that the consideration of snowline positions alone in the determination of the elemental ratios in forming gas-giant exoplanets corresponds {\em only} to the scenario that protoplanetary discs fully inherit all material directly from the molecular cloud, and further, that 
{\em no chemistry} occurs in the disc as planets are forming.  
On the other hand, in a more realistic protoplanetary disc in which chemistry has occured both en route into the disc and within the disc post formation, oxygen-rich conditions dominate the gas-phase material in the disc midplane.
In this work we will demonstrate that this can complicate the interpretation of the elemental ratios measured in hot Jupiter atmospheres when relating their potential formation locations to the dominant volatile snowlines in protoplanetary discs.

\section{Results}
\begin{figure*}
\includegraphics[width = 1.1\textwidth]{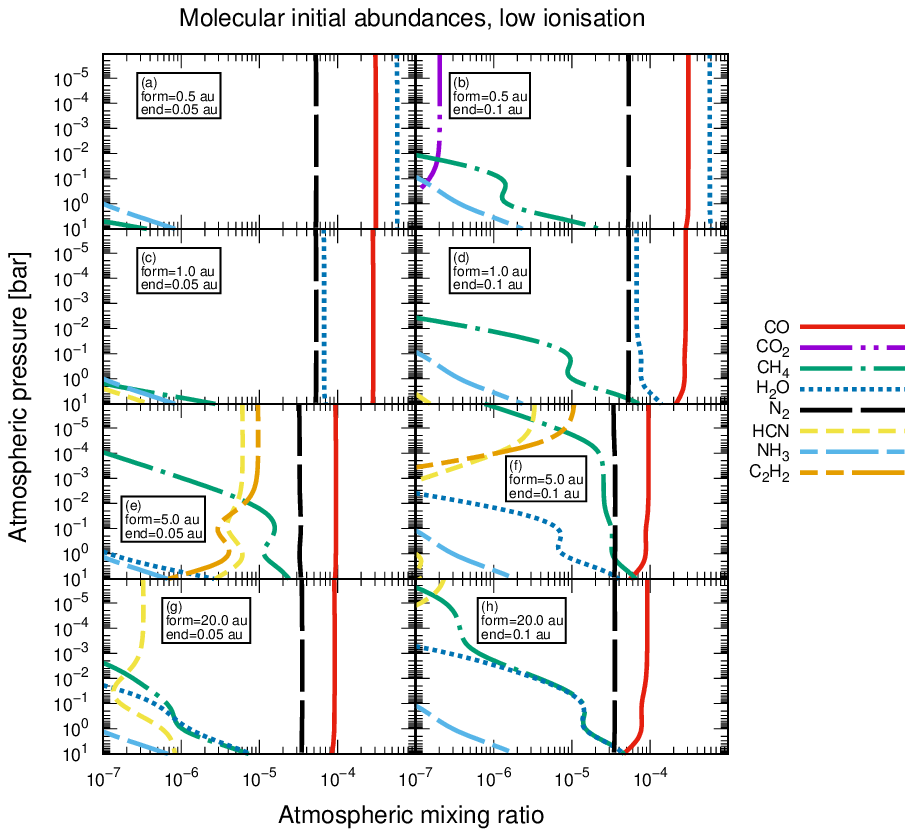}
\vspace{0.5cm}
\caption{Mixing ratios for major volatiles (\ce{CO}, \ce{CO2}, \ce{CH4}, \ce{H2O}, \ce{N2}, \ce{HCN}, \ce{NH3}, and \ce{C2H2}) in a hot Jupiter atmosphere, for eight different conditions for the formation and current position of the planet. 
In these panels, we assume that the midplane chemical evolution prior to atmospheric formation has taken place under low ionisation conditions and that the chemistry has begun with molecular initial abundances (the inheritance scenario).
The left-hand four panels (panels a, c, e, g) assume that the hot Jupiter has migrated to an orbital distance of $D=0.05$~au subsequent to formation of the atmosphere, whereas the right-hand four panels (panels b, d, f, h) assume that the planet has migrated to $D=0.1$~au. The results presented in the first (panels a and b), second (panels c and d), third (panels e and f), and fourth rows (panels d and h) assume assembly of the atmosphere at 0.5, 1.0, 5.0, and 20.0~au, respectively.}
\label{Figure3_new2}
\end{figure*}

\begin{figure*}
\includegraphics[width = 1.1\textwidth]{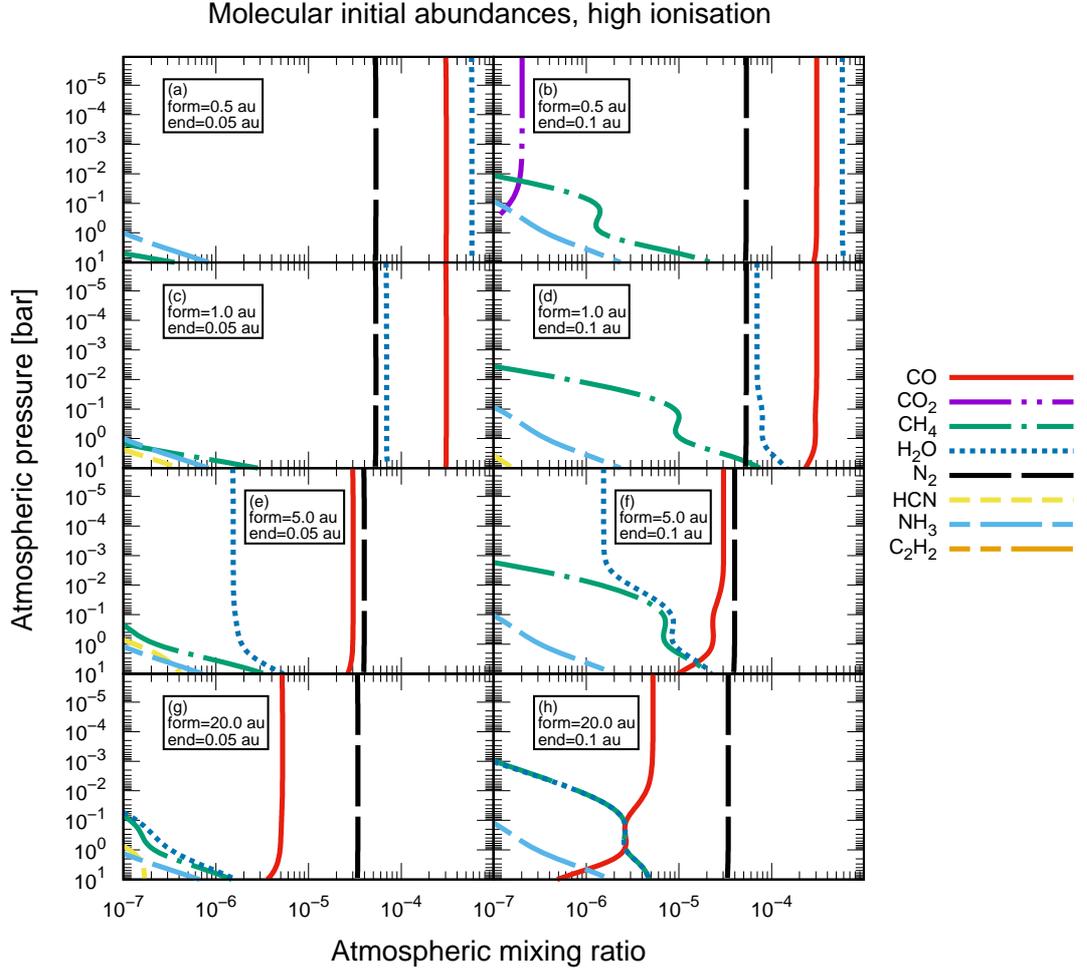}
\vspace{0.5cm}
\caption{The data presented are the same as described for Figure~\ref{Figure3_new2}, except for the case of the inheritance scenario (molecular initial abundances) and a high ionisation.}
\label{Figure4_new2}
\end{figure*}

\begin{figure*}
\includegraphics[width = 1.1\textwidth]{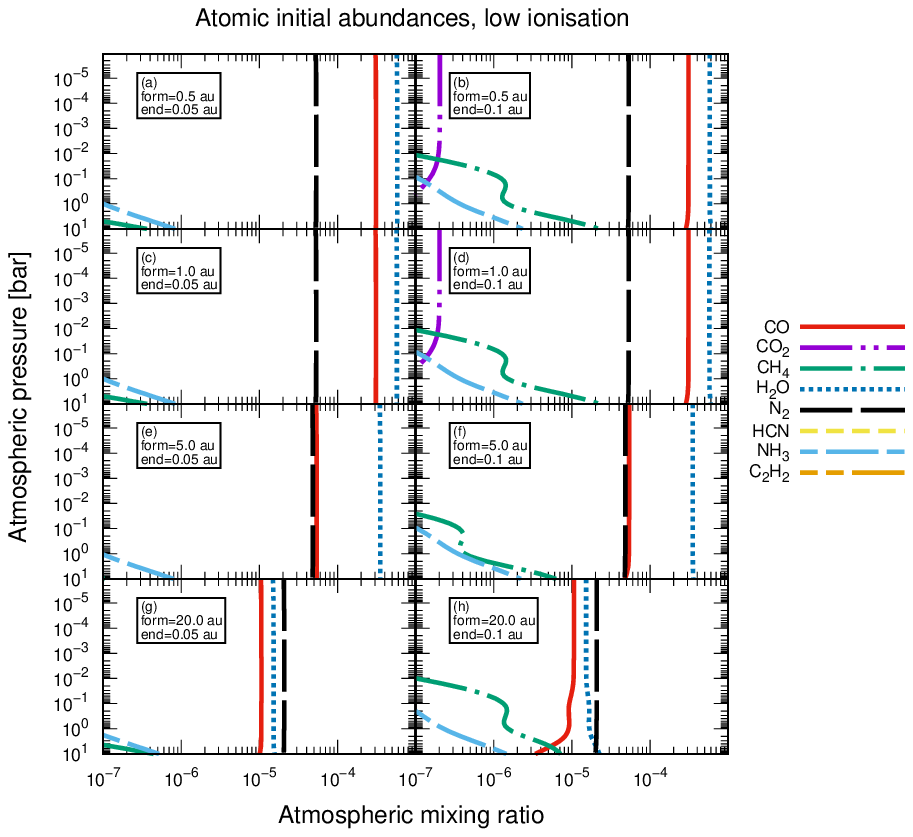}
\vspace{0.5cm}
\caption{The data presented are the same as described for Figure~\ref{Figure3_new2}, except for the case of the reset scenario (atomic initial abundances) and a low ionisation.}
\label{Figure5_new2}
\end{figure*}

\begin{figure*}
\includegraphics[width = 1.1\textwidth]{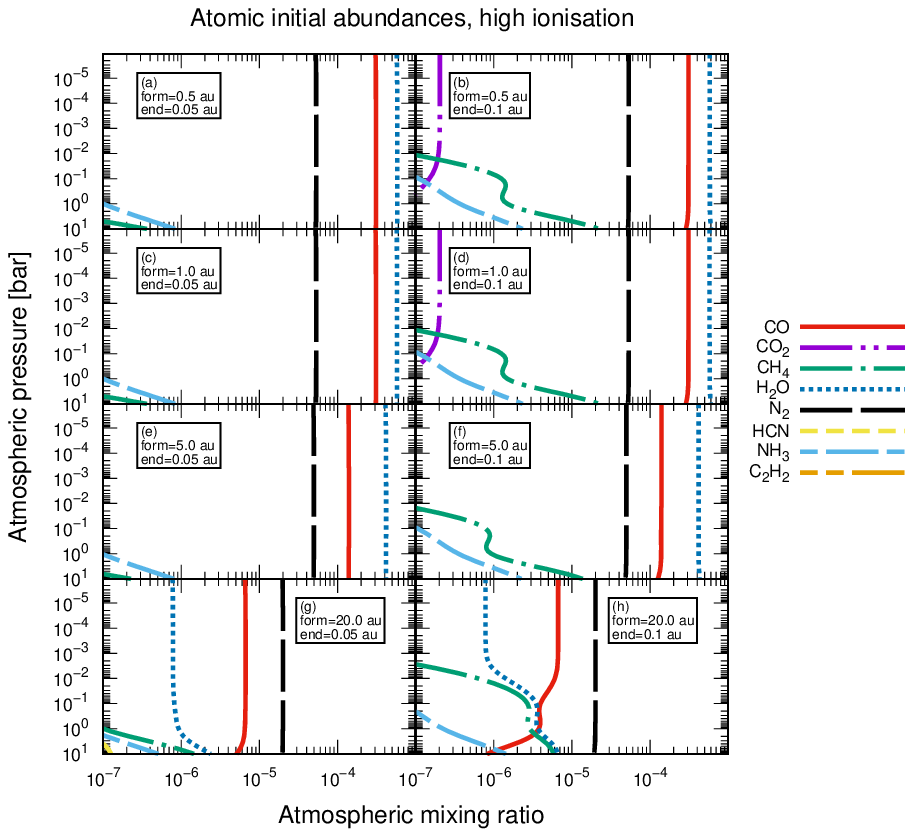}
\vspace{0.5cm}
\caption{The data presented are the same as described for Figure~\ref{Figure3_new2}, except for the case of the reset scenario (atomic initial abundances) and a high ionisation.}
\label{Figure6_new2}
\end{figure*}

The computed atmospheric mixing ratios for all model setups are presented in Figures \ref{Figure3_new2} to \ref{Figure6_new2}.
In each figure, mixing ratios for major volatiles (\ce{CO}, \ce{CO2}, \ce{CH4}, \ce{H2O}, \ce{N2}, \ce{HCN}, \ce{NH3}, and \ce{C2H2}) in a hot Jupiter atmosphere are shown, for eight different conditions for the formation location and current position of the gas-giant planet (stated in the inset box). 
We choose these volatiles because they are the most abundant carriers of C, O, and N, under chemical equilibrium conditions. 
We adopt the usual convention in which the $x$-axes show the mixing ratios (relative to the total gas-phase number density), and the $y$-axes show the atmospheric pressure in bars (increasing with depth into the atmosphere). 
As described in Section~2, four different initial elemental abundances were assumed for the 1 Myr disc midplane chemical evolution
corresponding to either the ``inheritance'' or the ``reset'' scenario, and the high or low ionisation level.  
This is indicated in each Figure.  
In addition to the presented Figures, in Table 2 we show the atmospheric CO, \ce{CH4}, and \ce{H2O} abundances and 
estimated C/O ratios at 0.5 bar (see also Section~4.1).

In the subsequent subsections, we describe the behavior of the mixing ratios of the dominant volatiles in the $0.1-1$ bar region because this is the pressure region most relevant for observations in Section 3.1. Following that, we describe the same for the less abundant volatiles in Section 3.2.

\subsection{Trends in atmospheric mixing ratios of major volatiles}

The variation in elemental ratios imposed by chemical evolution in the planet-forming regions of disc midplanes leads to a wide diversity of atmospheric compositions.
Looking first at the atmospheric mixing ratio profiles of CO (red solid lines in Figures~\ref{Figure3_new2} to \ref{Figure6_new2}), in general, those atmospheres assembled closer to the star, at 0.5 and 1~au (top two panels) are more abundant in CO ($\gtrsim 10^{-4}$), than those assembled farther out in the disc at 5~au and 20~au (bottom two panels; $10^{-7} - 10^{-4}$).  
CO is the most abundant volatile (excluding \ce{H2} and He) in several of the scenarios, in particular those in which the C/O ratio tends towards, or is greater than, $\approx 1$ (see Table~\ref{tab:1}).
Further, the mixing ratio of CO tends to remain constant with pressure throughout each atmosphere, except for those assembled beyond 5~au and that have migrated to 0.1~au (bottom right panels). In these atmospheres, the mixing ratio of CO decreases with pressure at pressures with $\gtrsim 0.1$~bar. 
This is because the atmospheric temperature at $\sim$1 bar for a gas giant at $D = 0.1$~au is lower ($T \approx 1500$~K) than that for a planet at $D = 0.05$~au ($T \approx 2100$~K). The CO and \ce{CH4} mixing ratios at $T \sim 1000-1500$~K are sensitive to both temperature and the C/O ratio (e.g., \citealt{Madhusudhan2012}).
 
There is a larger variation in the mixing ratio for \ce{H2O} (blue dotted lines in Figures~\ref{Figure3_new2} to \ref{Figure6_new2}), both across models, and within individual atmospheres, than found for CO, when the elemental abundance ratio satisfies C/O$>0.8$.
In the case of a hot Jupiter which has migrated to 0.1~au (right-hand panels), 
the planet 
has similar abundance or
more abundant in \ce{H2O} at $0.1 - 1$~bar than one which has migrated to 0.05~au (left-hand panels). 
For all planets for which the atmosphere is assembled at 0.5~au (top panels), water is the most abundant volatile (after \ce{H2} and He).  
For atmospheres assembled beyond 0.5~au (bottom three panels) within a disc with inherited abundances (Figures~\ref{Figure3_new2} and \ref{Figure4_new2}), the mixing ratio of \ce{H2O} decreases as the formation location radius of the planet is increased.  
This trend manifests because in the inheritance scenario, most oxygen is locked up in water ice beyond the water snowline 
so that the C/O ratio is $\gtrsim 0.8$ at 1~au and beyond.  
When the planet formation location is at 5 au or 20 au,
the \ce{H2O} mixing ratio is $\lesssim 10^{-5}$ at $0.1-1$~bar. 
In some cases, \ce{H2O} competes with CO as the main oxygen carrier and \ce{CH4} competes with CO as the main carbon carrier at high pressure region.
On the other hand, for the reset scenario (Figures~\ref{Figure5_new2} and \ref{Figure6_new2}), \ce{H2O} is the most abundant volatile for all atmospheres assembled within 20~au (top three panels) with a spread of a factor of only two in mixing ratios.  
This reflects the lower C/O ratio in these atmospheres compared with those considered in the inheritance scenario. 
This lower C/O ratio is because species other than the primary volatiles are produced in non-negligible quantities in the disc midplane in the case of chemical reset, including \ce{O2}, HCN, and NO. The high abundance of \ce{O2} in the gas phase ($\sim 10^{-4}$) within its snowline reduces the C/O ratio to $\lesssim 0.4$ \citep{Eistrup2016, Eistrup2018}. 
Under interstellar and circumstellar conditions, water ice formation occurs via hydrogenation of O and OH on dust grain surfaces \citep[see, e.g.,][]{Linnartz2015}: gas-phase formation of water is not efficient with the canonical abundance $\sim 10^{-4}$ reached only in the innermost hot disc midplane ($< 0.3$~au; see the discussion in \citealt{Eistrup2016}).  
In the reset scenario, gas-phase formation of \ce{O2} is able to capture the available atomic oxygen faster than the formation of water ice, because in the latter case the temperature of the dust grains between the \ce{H2O} and \ce{O2} snowlines is too high ($> 24$~K) for hydrogen to efficiently stick to dust-grain surfaces \citep{Walsh2015, Eistrup2016}.

For formation at 20~au in the reset scenario (bottom panels in Figures~\ref{Figure5_new2} and \ref{Figure6_new2}), the mixing ratio of water drops below $10^{-4}$, and \ce{N2} becomes the most abundant volatile (excluding \ce{H2} and He).  
For the case of a low ionisation level in the disc midplane (Figure~\ref{Figure5_new2}), water remains the primary O-bearing species in the atmosphere, 
again reflecting the low C/O ratio in this case.  
However, in the 
case of a high ionisation level in the disc midplane (Figure~\ref{Figure6_new2}), the C/O ratio tends toward 1 and CO becomes the main C- and O-bearing species at $\sim0.1-1$~bar. 

The atmospheric mixing ratios for \ce{CH4} (dot-dashed green lines in Figures~\ref{Figure3_new2} to \ref{Figure6_new2}) are generally lower at pressures of $0.1-1$~bar when compared with those for \ce{H2O} and CO. 
The exception to this is the case of formation at 5~au in the inheritance scenario and at a low ionisation (see panels e and f in Figure~\ref{Figure3_new2}) where the mixing ratio of \ce{CH4} exceeds that of water. This location corresponds to formation between the \ce{CO2} and \ce{CH4} snowlines.
This scenario generates the most carbon rich conditions (C/O ratio $ = 1.27$ at 5~au; see Table~\ref{tab:1}).
For this scenario and for a planet that ends up at 0.1~au, \ce{CH4} 
has a mixing ratio of $>10^{-5}$ at $>10^{-4}$~bar.
For a planet currently at 0.05~au for this case, \ce{CH4} maintains the same mixing ratio at $>10^{-2}$~bar.  
A similar profile is seen for a planet with a formation location of 20 au, and a final orbital radius of 0.1 au (panel h in Figure \ref{Figure3_new2}).
In all other cases, the mixing ratio of \ce{CH4} has a steep gradient with increasing pressure, possessing a negligible value ($\ll 10^{-7}$) at $<10^{-3}$ bar.
For planets that are currently at 0.05~au and that were formed in a disc with atomic initial conditions and/or a high ionisation (Figures~\ref{Figure4_new2} to \ref{Figure6_new2}) 
the mixing ratio of \ce{CH4} reaches $> 10^{-6}$ only at the highest pressures considered here, $> 1$ bar.
In planets currently at 0.1~au in all scenarios
except atomic initial abundances with formation location of 5 au (panel f of Figures~\ref{Figure5_new2} and \ref{Figure6_new2}) where both C/H and C/O ratios are low,
there is a layer of \ce{CH4} with a mixing ratio of $>10^{-6}$ between 0.1 and 1~bar.
This is in accordance with model results from \citet{Madhusudhan2012} and \citet{Moses2013}, who have shown that \ce{CH4} is efficiently formed in cooler atmospheres ($\lesssim 1500$~K) and/or with a high C/O ratio ($>$1.0). 

Another carbon-bearing molecule with an appreciable mixing ratio ($>10^{-6}$) in some of the modeled atmospheres is \ce{C2H2} (orange dashed lines in Figures~\ref{Figure3_new2} to Figure~\ref{Figure6_new2}), albeit only for the inheritance scenario, in which a low ionisation is assumed, and for which the planet has formed at 5.0 au (see panels e and f in Figure \ref{Figure3_new2}). 
For both final orbital radii in these scenarios, \ce{C2H2} reaches a peak mixing ratio of $\sim 10^{-5}$ at lower atmospheric pressures. 
For the planet currently at 0.05~au,  the mixing ratio remains fairly constant at this level up to a pressure of $\sim 10^{-3}$~bar, and reducing to a value of a few times $10^{-6}$ at $1 - 10$ bar.
Under chemical equilibrium conditions, efficient formation of \ce{C2H2} is achieved only at higher C/O ratios ($> 1.0$) and higher temperatures ($T>2000$~K) \citep{Madhusudhan2012}.

With regards to the major nitrogen-bearing species, mixing ratios for \ce{N2} and HCN are also shown in Figures~\ref{Figure3_new2} to Figure~\ref{Figure6_new2} (black dashed lines and yellow dashed lines, respectively). 
For HCN, the mixing ratio only reaches a significant value ($>10^{-6}$) for the inheritance scenario at low ionisation for a formation location of 5~au (panels e and f of Figure~\ref{Figure3_new2}).
The mixing ratio reaches $>10^{-6}$ up to 10 bar for a final orbit of 0.05~au, and only at $<10^{-4}$~bar for a final orbit of 0.1 au.
This is because it is only for this scenario that the C/O ratio exceeds 1 (see Table \ref{tab:1}), and as seen in previous works, the HCN abundance under chemical equilibrium conditions is very dependent on the C/O ratio (e.g., \citealt{Madhusudhan2012, Moses2013, Hobbs2019}). Thus the situation for HCN is similar to that discussed for \ce{C2H2}.

For \ce{N2}, the maximum atmospheric mixing ratio across all panels is $\sim 5 \times 10^{-5}$. For formation locations at 5 au and 20 au, the mixing ratios are lower, by at most a factor of 4 lower than the maximum mixing ratio seen across the models. As shown in Table \ref{tab:1}, Figure~\ref{Figure2_new2}, and \citet{Eistrup2016}, this small dispersion in the mixing ratios for \ce{N2} is because the N/H ratio through the disc midplane shows the least variability with radius. This is because the main carrier of gas-phase nitrogen in the disc is \ce{N2}, and the snowline for \ce{N2} lies beyond 30~au in the model considered in \citet{Eistrup2016}. 
\ce{N2} decreases at 5 au and 20 au by a factor of a few because when beginning the disc chemistry, other nitrogen carriers, including HCN and NO, are able to form from the available free atomic nitrogen. Both of HCN and NO freeze out at higher temperatures than \ce{N2} thus removing gas-phase nitrogen from the outer disc. The mixing ratios for \ce{N2} for each scenario remain constant with altitude, as found in other works \citep[e.g.,][]{Venot2012,Moses2013}, and this is because of its strong intramolecular bond. 

\subsection{Trends in atmospheric mixing ratios of minor volatiles}
Mixing ratio profiles for the more minor atmospheric constituents, \ce{CO2} and \ce{NH3}, are also shown in Figures~\ref{Figure3_new2} to \ref{Figure6_new2} (purple dotted lines and blue dashed lines, respectively). 
We consider these species to be minor because their mixing ratios typically do not reach $\gtrsim 10^{-6}$ in the atmosphere in any scenario considered here. 
Nonetheless, given that both species are detected in cool brown dwarfs \citep[e.g.,][]{Line2017}, and that both species are sensitive to departures from 
chemical equilibrium \citep[][]{Moses2011,Moses2013,Venot2012, Hobbs2019}, it is worth to discuss their behavior here.

For \ce{CO2}, the maximum atmospheric mixing ratios across all scenarios is $2\times10^{-7}$. 
\ce{CO2} does not reach $\gtrsim 10^{-7}$ in any planets that have migrated to 0.05~au.
For the inheritance scenario (shown in Figures~\ref{Figure3_new2} and \ref{Figure4_new2}), the \ce{CO2} mixing ratio only reaches this value for a planet formed within the \ce{H2O} snowline and which has migrated to 0.1~au.  
Further this is only reached high in the atmosphere at pressures below $0.01$~bar. 
For all other planet formation locations and current locations, the mixing ratio for \ce{CO2} is $\lesssim 10^{-7}$.  
This is because the abundance of oxygen available in the gas phase falls by a factor of a few between the \ce{H2O} and \ce{CO2} snowlines, and by more than an order of magnitude beyond the \ce{CO2} snowline (see Table \ref{tab:1} and \citealt{Eistrup2016}).  

For those planets that have migrated to 0.1~au, a similar behaviour is found in the reset scenario (shown in Figures~\ref{Figure5_new2} and \ref{Figure6_new2}), except that \ce{CO2} is also present at a level $\sim 10^{-7}$ in the planets that formed at 1~au, that is, between the \ce{H2O} and \ce{CO2} snowlines.  
This is because when beginning the disc chemistry with atomic initial abundances, more volatile carriers of oxygen are formed (e.g., \ce{O2} and \ce{CO}), such that the abundance of gas-phase oxygen does not vary across the \ce{H2O} snowline \citep{Eistrup2016}.  
Further, the drop in oxygen abundance across the \ce{CO2} snowline is only a factor of a few, which is contrasted with the order of magnitude decrease seen in the inheritance scenario.

\ce{NH3} reaches appreciable mixing ratios $\gtrsim 10^{-7}$ deep in the atmosphere only, typically $\gtrsim 0.01$~bar. 
Across all scenarios, the mixing ratio then increases with depth, reaching values $\sim 10^{-6}$ at 10~bar. 
There is little difference in the behaviour of \ce{NH3} across the different scenarios, except that those planets that have migrated to 0.1~au have a few times higher mixing ratios than those that have migrated to 0.05~au.  

In our calculations, we also include \ce{C2H4} as an atmospheric constituent. For \ce{C2H4}, the behaviour of the mixing ratios closely follow that of \ce{C2H2}; however, the values are lower by around three orders of magnitude.
Therefore, we did not include the result in Figures~\ref{Figure3_new2} to \ref{Figure6_new2}.

\section{Discussions}

\subsection{What are the key molecules to observationally estimate the C/O ratios of gas-giant atmospheres?}

In Section 3 we reported that the mixing ratios of CO and \ce{N2} in each atmospheric model are relatively insensitive to the initial conditions adopted in the protoplanetary disc model that sets the gas-phase elemental ratio in the atmosphere.
In contrast, the other considered atmospheric volatiles, in particular, \ce{H2O}, \ce{CH4}, \ce{C2H2}, and \ce{HCN}, are very sensitive to the C/O ratio.
As the C/O ratio tends towards 1, the mixing ratio of \ce{CH4} increases in the lower atmosphere, and when the C/O ratio exceeds 1, those of \ce{C2H2} and \ce{HCN} also increase in the upper atmosphere.  
On the other hand, the mixing ratios of O-bearing species (\ce{H2O} and \ce{CO2}) decrease as the C/O ratio increases.
These results are consistent with previous studies, which have investigated the atmospheric chemical structure using a range of C/O ratios 
\citep[e.g.,][]{Madhusudhan2012, Moses2013, Mordasini2016, Drummond2019}.

However, the goal of this work was to assess the role of chemistry in protoplanetary disc midplanes on setting the elemental composition of a gas-giant atmosphere. 
In particular, we wish to assess how elemental change across snowlines are impacted by chemistry, and whether or not the measurement of elemental ratios in a gas-giant atmosphere can pinpoint the formation location of a close-in gas-giant planet.   
In this subsection we discuss first the accuracy of determining the C/O ratios from observations of exoplanetary atmospheres using the currently observable volatiles.
Here, we determine the C/O ratios of each atmospheric model using the mixing ratios of the major carbon- and oxygen-bearing molecules only (\ce{H2O}, CO, and \ce{CH4}).
Table~\ref{tab:2} shows the atmospheric \ce{H2O}, CO, and \ce{CH4} abundances and estimated C/O ratios at 0.5~bar. 

Comparing the values in Tables~\ref{tab:1} and \ref{tab:2}, in most cases the C/O ratios estimated from observations of \ce{H2O}, CO, and \ce{CH4} only, reproduce the values from the initial disc models with a precison of $\sim0.01$, well within the error bars anticipated for atmospheric retrieval from future observations of gas-giant exoplanets \citep[e.g.,][]{Greene2016,Schlawin2018,Changeat2020}.
The exception to this is for our most carbon-rich case (the inheritance scenario and a low ionisation) and formation at 5~au and migration to 0.05~au.
In that case, the difference between the input C/O ratio and ``measured" C/O ratio is 0.14. 
This is due to the neglect of \ce{C2H2} and \ce{HCN} as carriers of carbon in the measurement of the C/O ratio in the atmosphere (see Figure~\ref{Figure3_new2}).  
Thus for atmospheres with super-solar C/O ratios, observations of \ce{HCN} and \ce{C2H2} are needed to estimate the C/O ratios precisely, 
especially if the temperature of the planet is high (e.g., $> 2000$~K at $0.1-1$~bar). 
It should be noted here that our work predicts that carbon-rich hot Jupiter planets are likely to be rare. 
This is due to the very particular circumstances under which carbon-rich conditions arise in disc midplanes, 
namely assuming full inheritance of abundances from the molecular cloud and that no chemistry occurs during disc formation and evolution.

\ce{H2O}, CO, and \ce{CH4} have strong features at near- to mid-infrared wavelengths, and have been detected in recent observations in mainly bright hot Jupiters 
(e.g., \citealt{Snellen2010, Kreidberg2014, Brogi2016, Birkby2017, Line2017, Samland2017, Hawker2018, Guilluy2019, Madhusudhan2019}).
In addition, HCN has recently been detected for bright hot Jupiters \citep{MacDonald2017, Hawker2018, Cabot2019}.
Using next-generation facilities, such as JWST, ARIEL, SPICA, and future ground-based telescopes (e.g., E-ELT and TMT), it is anticipated that C/O ratios will be observationally determined with much higher precision than currently possible for many hot Jupiters (e.g., \citealt{Greene2016, Greene2019, Schlawin2018, Tinetti2018, Venot2018, Bowler2019, Brogi2019, Madhusudhan2019, Changeat2020, Venot2020}). 

\begin{table*}
	\centering
	\caption{The atmospheric CO, \ce{CH4} and \ce{H2O} mixing ratios and estimated C/O ratios at 0.5 bar.}
	\label{tab:2}
	\begin{tabular}{lcccccr}
                 \hline		
                 \multicolumn{6}{c}{Molecular initial abundances} \\
		 \hline
		 & Formation & \ce{CO} & \ce{CH4} & \ce{H2O} & C/O\\
		\hline
		&0.5 au &$3.08\times10^{-4}$&$4.46\times10^{-9}$&$5.79\times10^{-4}$&0.35 \\
		low&1 au &$2.86\times10^{-4}$&$3.56\times10^{-8}$&$6.73\times10^{-5}$&0.81\\
		ionisation&5 au &$9.50\times10^{-5}$&$1.21\times10^{-5}$&$6.62\times10^{-8}$&{\bf 1.13}$^\ast$\\ 
                 end=0.05au&20 au&$9.08\times10^{-5}$&$8.45\times10^{-7}$&$9.04\times10^{-7}$&1.00\\ 
		&&  &  & &\\	
		&0.5 au &$3.07\times10^{-4}$&$1.28\times10^{-6}$&$5.81\times10^{-4}$&0.35 \\
		low&1 au &$2.77\times10^{-4}$&$8.80\times10^{-6}$&$7.61\times10^{-5}$&0.81\\
		ionisation&5 au &$8.85\times10^{-5}$&$3.21\times10^{-5}$&$6.68\times10^{-6}$&1.27\\
                 end=0.1 au&20 au&$7.82\times10^{-5}$&$1.39\times10^{-5}$&$1.36\times10^{-5}$&1.00\\
 		&&  &  & &\\	                
                 &0.5 au &$3.06\times10^{-4}$&$4.43\times10^{-9}$&$5.79\times10^{-4}$&0.35\\
		high&1 au &$3.06\times10^{-4}$&$3.70\times10^{-8}$&$6.95\times10^{-5}$&0.82\\
		ionisation&5 au &$3.01\times10^{-5}$&$1.44\times10^{-7}$&$1.75\times10^{-6}$&0.95\\
                 end=0.05 au&20 au&$4.99\times10^{-6}$&$1.71\times10^{-7}$&$2.45\times10^{-7}$&0.99\\
		&&  &  & &\\	
		&0.5 au &$3.05\times10^{-4}$&$1.27\times10^{-6}$&$5.81\times10^{-4}$&0.35\\
		high&1 au &$2.98\times10^{-4}$&$9.14\times10^{-6}$&$7.87\times10^{-5}$&0.82\\
		ionisation&5 au &$2.35\times10^{-5}$&$6.81\times10^{-6}$&$8.36\times10^{-6}$&0.95\\
                 end=0.1 au&20 au&$2.69\times10^{-6}$&$2.55\times10^{-6}$&$2.55\times10^{-6}$&1.00\\ 
		\hline		
                \multicolumn{6}{c}{Atomic initial abundances} \\
		\hline
		 & Formation & CO & \ce{CH4} & \ce{H2O} & C/O \\
		\hline		
                 &0.5 au &$3.10\times10^{-4}$&$4.48\times10^{-9}$&$5.79\times10^{-4}$&0.35\\ 
		low&1 au &$3.09\times10^{-4}$&$4.53\times10^{-9}$&$5.73\times10^{-4}$&0.35\\
		ionisation&5 au &$5.39\times10^{-5}$&$1.29\times10^{-9}$&$3.51\times10^{-4}$&0.13\\
                 end=0.05 au&20 au&$1.06\times10^{-5}$&$5.87\times10^{-9}$&$1.51\times10^{-5}$&0.41\\
		&&  &  & &\\	
		&0.5 au &$3.09\times10^{-4}$&$1.28\times10^{-6}$&$5.81\times10^{-4}$&0.35\\ 
		low&1 au &$3.08\times10^{-4}$&$1.30\times10^{-6}$&$5.75\times10^{-4}$&0.35\\ 
		ionisation&5 au &$5.36\times10^{-5}$&$3.69\times10^{-7}$&$3.51\times10^{-4}$&0.13\\
                 end=0.1 au&20 au&$9.23\times10^{-6}$&$1.35\times10^{-6}$&$1.65\times10^{-5}$&0.41\\
 		&&  &  & &\\	                
		&0.5 au &$3.10\times10^{-4}$&$4.48\times10^{-9}$&$5.79\times10^{-4}$&0.35\\ 
		high&1 au &$3.10\times10^{-4}$&$4.50\times10^{-9}$&$5.77\times10^{-4}$&0.35\\
		ionisation&5 au &$1.39\times10^{-4}$&$2.81\times10^{-9}$&$4.16\times10^{-4}$&0.25\\
                 end=0.05 au&20 au&$6.57\times10^{-6}$&$6.34\times10^{-8}$&$8.71\times10^{-7}$&0.89\\
		&&  &  & &\\			
		&0.5 au &$3.09\times10^{-4}$&$1.28\times10^{-6}$&$5.81\times10^{-4}$&0.35\\
		high&1 au &$3.09\times10^{-4}$&$1.29\times10^{-6}$&$5.78\times10^{-4}$&0.35\\
		ionisation&5 au &$1.39\times10^{-4}$&$8.03\times10^{-7}$&$4.17\times10^{-4}$&0.25\\
                end=0.1 au&20 au&$3.94\times10^{-6}$&$2.72\times10^{-6}$&$3.51\times10^{-6}$&0.89\\
		\hline		
\multicolumn{6}{l}{$^\ast$Highlighted is the only case for which the measured C/O ratio from \ce{H2O}, CO,} \\
\multicolumn{6}{l}{and \ce{CH4} deviates by more than 0.01 from the input value.}
	\end{tabular}
\end{table*}

\subsection{Chemical imprints of snowlines in gas-giant atmospheres}
In many previous studies, it was assumed that the C/O ratio in disc midplane gas generally approaches or exceeds 1 outside the water snowline \citep[see, e.g.,][]{Oberg2011}.
However, according to our calculations such high gas-phase C/O ratios are achieved only beyond the \ce{CO2} snowline in a disc with a low ionization (i.e., minimal chemical evolution) and molecular initial abundances (i.e., the inheritance scenario).  
We find that only in this case does the C/O ratio exceed 1, and only between the \ce{CO2} and \ce{CH4} snowlines.
Moreover, for a disc with inherited abundances and with a high level of ionisation, the gas-phase C/O ratio remains less than 1 always within the CO snowline.    
For the reset scenario, the maximum C/O ratio achieved in the gas is $\approx 0.4$.  
We note here that in the above, we are assuming that the disc is $1$~Myr old, and that this coincides with the epoch of planet formation; \citet{Eistrup2018} show that as the disc further evolves towards $\approx 7$~Myr for the case of a high ionisation, the gas-phase C/O ratio tends towards 0.4 within the CO snowline, i.e., the gas becomes increasingly oxygen rich with time.  
Thus we make a further prediction that the existence of carbon-rich gas-giant planet atmospheres via accretion of disc midplane gas is only possible if the planet forms very early in the disc lifetime, before significant chemical evolution can take place.  
We note here that for a Jupiter-mass planet, the slow envelope accretion phase is estimated to last $\approx 0.5$ to a few~Myr, whereas the runaway accretion phase, in which the planet accretes most of its envelope, lasts $\approx 10^{4} - 10^{6}$~yr \citep[e.g.,][]{Helled2014PPVI}.

According to Table~\ref{tab:1} and Figure~\ref{Figure2_new2} (see also \citealt{Eistrup2016}), in the case of the inheritance scenario, the disc oxygen abundance in the gas phase between the \ce{CO2} and \ce{CH4} snowlines ($\lesssim6\times10^{-5}$) is much lower than the oxygen abundance in the same location in the reset scenario ($\gtrsim3\times10^{-4}$). 
The former value is roughly one order of magnitude lower than the value in the solar atmosphere ($4.9\times10^{-4}$; \citealt{Asplund2009}). 
This is because in the reset scenario water ice is not efficiently formed in the outer disc leading to oxygen-rich gas (see Section~3 and the discussions in \citealt{Eistrup2016}).
In addition, outside the \ce{CO2} snowline, the disc carbon abundance in the gas phase is lower in the high ionisation case compared with the low ionisation case, and the C/O ratio is less than 1 everywhere in the former case. 
This is because of destruction of \ce{CH4} gas and the production of \ce{O2} gas \citep{Eistrup2016}.
In contrast, the abundances of nitrogen-bearing species in the gas phase are relatively constant across all cases. 
Therefore, it is difficult to constrain the planet formation locations from nitrogen-bearing molecules, although they are important for determining the overall metallicities.

Summarizing the discussions above, if a hot Jupiter atmosphere has a high C/O ratio ($> 1$) and a sub-stellar oxygen abundance, we predict that they will have formed between the \ce{CO2} and \ce{CH4} snowlines in discs in which negligible chemical evolution has taken place and which have inherited molecular abundances from the parent molecular cloud.
This constitutes one of the only cases where the C/O ratio and the O/H ratio together can uniquely identify a formation location of a gas-giant planet relative to the positions of snowlines in the protoplanetary disc, in addition to the nature of the disc in which the planet has formed.
Considering again the elemental ratios listed in Table~\ref{tab:1}, there exist several other scenarios in which a gas-giant planet's formation location could be identified from spectroscopy of its atmosphere.    
First, in the case of a C/O ratio $\approx 0.8 - 1$, for a close-to-solar O/H ratio ($\sim 10^{-4}$) this corresponds to formation between the \ce{H2O} and \ce{CO2} snowlines, in a disc which has inherited its molecular abundances from its parent cloud; however, the degree of chemical processing within the disc cannot be constrained from such a planet.
Secondly in the case of a similar C/O ratio but a sub-solar O/H ratio, if O/H $\sim 10^{-5}$ then we have several degenerate scenarios: either the planet formed beyond the \ce{CH4} snowline in a disc with inherited abundances and a low ionisation, or it formed between the \ce{CO2} and \ce{CH4} snowlines in a disc with inherited abundances and a high ionisation, or it formed beyond the \ce{CH4} snowline in a disc which has experienced chemical reset and which has a high level of ionisation.
For this case, we can only conclude that the planet formed beyond the \ce{CO2} snowline.  
For the same C/O ratio and an O/H ratio $\sim 10^{-6}$, then this also uniquely corresponds to formation beyond the \ce{CH4} snowline in a disc with inherited abundances and a high level of ionisation.   

For all other scenarios (see Tables~\ref{tab:1} and \ref{tab:2}) the C/O ratio is $\lesssim 0.4$.  
In the traditional picture \citep{Oberg2011}, this measurement would imply that the planet uniquely formed within the \ce{H2O} snowline.  
However, our results show that this interpretation is not so clear-cut.  
For a value of the C/O ratio of $\approx 0.35 - 0.4$ and a solar-like O/H ratio ($\sim 10^{-4}$), it is also possible that the planet formed between the \ce{H2O} and \ce{CO2} snowlines in a disc in which chemical reset has taken place.  
Thus, for this combination of C/O and O/H it is possible to constrain the planet's formation location to within the \ce{CO2} snowline only.  
For a lower O/H $\sim 10^{-5}$, we predict that the planet has formed between the \ce{CH4} and CO snowlines in a disc in which chemical reset has taken place and in which there was a low level of ionisation.  
The reset scenario also generates a further unique scenario, which is a sub-solar C/O ratio ($\lesssim 0.35$).  
Such a measurement, in conjunction with a close to solar O/H ratio ($\sim 10^{-4}$), would point to formation within the \ce{CO2} and \ce{CH4} snowlines in a disc in which chemical reset has taken place.  
This demonstrates the importance of retrieval tools to allow for the possibility of sub-solar C/O ratios.  

\begin{figure*}
\includegraphics[width=0.8\textwidth]{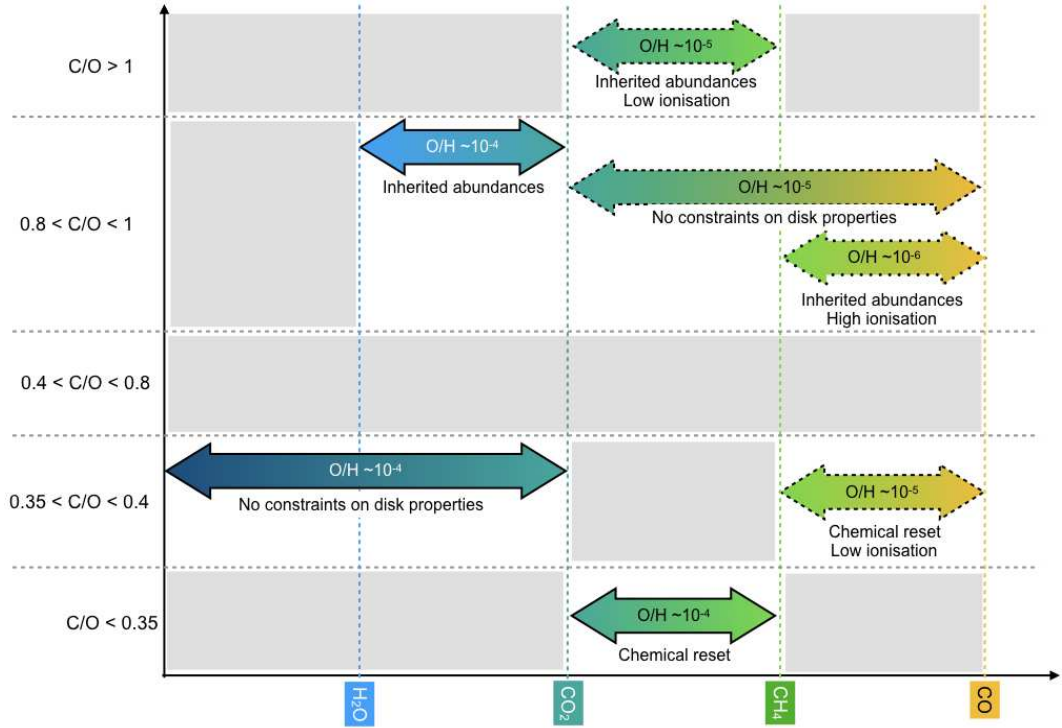}
\vspace{0.2cm}
\caption{Diagram showing the possible combinations of C/O and O/H ratios predicted by the chemical model. Note that both axes have arbitrary scales.  
The arrows highlight the possible range of formation locations of a gas-giant planet relative to the stated snowline locations for each possible combination of C/O (range indictated on the y axes) and O/H. The order of magnitude value for O/H is stated inset of the arrow, and is also indicated by the border line type: O/H~$\sim 10^{-4}$ (solid), $\sim 10^{-5}$ (dashed), and $\sim 10^{-6}$ (dotted).  
The possible disc properties which can be constrained from these combinations of C/O and O/H are listed below each arrow.
The greyed out regions are those for which the range of chemical models explored predict no possible solutions.}
\label{Figure7_new2}
\end{figure*}

The various scenarios described above are presented in diagram form in Figure~\ref{Figure7_new2}.
The consideration of chemical evolution in disc midplanes increases the number of possible combinations of C/O and O/H ratios above that traditionally considered by the exoplanet community.  
Despite several degeneracies arising in the extraction of possible disc properties, there exist unique combinations of C/O and O/H ratios
that reveal the formation location of the gas-giant planet relative to snowline locations and which allow some constraints on the properties of the discs within which those atmospheres have assembled.

\subsection{Pollution of atmospheres by solid accretion}
Protoplanetary discs are globally composed of $\approx 99\%$ gas and $\approx 1\%$ solids (in the form of refactory dust grains) by mass. 
Beyond snowlines, volatiles are depleted from the gas phase as ice on the surfaces of dust grains.  
In our atmospheric calculations, we assume that the planet has accreted its atmosphere solely from the disc midplane gas, that is, we do not include any contribution from icy dust grains and pebbles which may be accreted onto the planet during and/or after gas acquisition.
Such solid icy bodies impacting on planets can alter the metallicities of planetary atmospheres, and the solid bodies themselves may also experience effects such as grain growth, fragmentation, and radial drift prior to impact, all of which can change the C/O ratios of the ices that they host over time \citep[see, e.g.,][]{Moses2013, Mordasini2016}.
According to previous studies \citep[e.g.,][]{Oberg2011, Mordasini2016, Cridland2019a}, the accretion of such icy dust grains will act to reduce the C/O ratio in the atmosphere because the ices are oxygen rich and carbon depleted.
In the case of discs with abundances inherited from their parent molecular cloud, the C/O ratio of the ice is consistently $< 0.4$ inside the CO snowline.
In the reset scenario, the C/O ratio in the ice is consistenly $< 0.6$ inside the CO snowline.
Thus, if we consider icy component accretion post gas acquisition, the C/O ratio in the gas-giant atmosphere can only decrease.

Recent observations \citep[e.g.,][]{Madhusudhan2011, Moses2013, Brewer2017} suggest that some hot Jupiters may have super-stellar C/O ratios and sub-stellar O/H ratios which, based on our models, is only possible if the bulk of the planet's atmosphere were assembled from the gas-phase only.
Here we note that according to \citet{Kreidberg2015}, the values of molecular abundances and C/O ratios retrieved from data can vary greatly because they are strongly dependent on assumptions that are adopted in the retrieval models and the observational method (i.e., transmission spectra versus dayside emission spectra). 
They recommend that obtaining high-precision data with multiple observing techniques and 
phase-resolved emission with 3D atmospheric circulation modeling are both critical to obtain precise constraints on the chemistry and physics of exoplanetary atmospheres (see also the recent review of retrieval methods by \citealt{Barstow2020}).

In our calculations, we also do not consider the impact of core dissolution on the atmospheric composition.
\citet{Oberg2019} have argued that the core of Jupiter in solar system could have formed outside the \ce{N2} snowline ($>$30 au), and some elements (such as carbon and nitrogen) in Jupiter's gaseous envelope were subsequently enriched by core dissolution (see also \citealt{Bosman2019}). 
This seems in contrast with a sub-solar oxygen abundance recorded by observations with the Galileo probe;  
however, this measurement is generally assumed to not be representative of Jupiter's true composition \citep{Helled2014}. 
The current Juno mission is anticipated to finally reveal Jupiter's global oxygen abundance \citep{Li2020}. 
As discussed in \citet{Oberg2019} the elemental abundances of solids beyond the \ce{N2} snowline is expected to close to the solar value. 
Assuming core formation in the outer disc followed by gas accretion in the inner disc in conjunction with core dissolution, provides a mechanism to achieve super-solar abundances in gas-giant atmospheres because this formation scenario has effectively transported heavy elements from the outer disc inwards.

Remaining on the topic of transport of solids, we note here that recent models of disc midplane composition exploring the impact of radial drift of pebbles on the chemical abundances of the gas show C/O~$>> 1$ and super-solar elemental abundances between the \ce{CO2} and \ce{CH4} snowlines \citep{Booth2017, Booth2019}, whether or not chemistry is considered, and for viscous discs ($\alpha \gtrsim 10^{-3}$).
The C/O ratios and elemental abundances of the gas in discs can be significantly increased particularly around snowlines due to
the efficient sublimation of volatile ices hosted on pebbles that have been transported inwards \citep{Booth2017, Booth2019}.
Thus, this provides an additional mechanism for giant planets to acquire super-solar elemental abundances in their atmospheres with C/H$>10^{-4}$ (and O/H$>10^{-4}$ in some cases), and C/O~$>1$, if they undergo runaway accretion between the \ce{CO2} and \ce{CH4} snowlines.
As we discuss in the following paragraphs, our work also predicts that super-solar C/H and C/O ratios are only possible between the \ce{CO2} and \ce{CH4} snowlines with the added constraints. This implies fully inherited interstellar abundances as well as a low ionisation rate (negligible chemistry) within the disc midplane ($\lesssim 10^{-18}$~s$^{-1}$).

In the models presented in \citet{Booth2019} that include both chemistry and pebble accretion, it was concluded that material transport is faster than chemical modification even when assuming a high cosmic-ray ionisation rate of $\sim 10^{-17}$~s$^{-1}$.  
However, the chemical network used in \citet{Booth2019} is more simple than that adopted in \citet{Eistrup2016}, and neglected many grain-surface processes such as radical-radical recombination and cosmic-ray-induced processing of the ice \citep[see, e.g.,][]{Cuppen2017}.  
Including a more complex chemistry in a model with pebble drift would enable a comparison between the ice chemical processing timescale and the pebble drift timescale.
This would then determine whether or not the former is sufficiently long to preserve the composition of the icy pebbles during transport through the disc midplane.

We note that the models of \citet{Booth2019} also consider 
that the abundances in the molecular clouds have been fully inherited as initial molecular abundances of the protoplanetary disc.
Not yet investigated in conjunction with pebble drift is the scenario that some chemical modification has occurred during disc formation and early evolution, 
the most extreme assumption of which is that the disc material has undergone full chemical reset due to exposure to, e.g., shocks. 
For that scenario, \citet{Eistrup2016} and \citet{Eistrup2018} find C/O~$<< 1$ because the disc gas is depleted in \ce{CH4} and enhanced in \ce{O2}.  
Chemical reset at the earlier phases of disc evolution has not yet to be considered in disc midplane models with pebble drift.  Such simulations would test the robustness of predictions of super-solar C/O and C/H abundances between the \ce{CO2} and \ce{CH4} snowlines that are expected to not hold under the reset scenario.
Indeed, there is growing evidence that planet-building material in discs undergoes an earlier phase of chemical evolution at elevated temperatures relative to those found in the protoplanetary disc phase. For example, recent state-of-the-art observations of young discs with ALMA have shown that these discs are relatively warmer than their more evolved counterparts \citep[e.g.,][]{Lin2020,Vanthoff2020}.
%

The C/O ratio and bulk atmospheric elemental abundances provide important clues regarding the formation and evolution of gas giant planets.
Future studies will include contributions to gas-giant atmosphere from solid impactors and pebbles to quantify the influence on the elemental composition of the atmosphere and to investigate the potential for identifying signatures of past solid accretion and sublimation of the volatiles from pebbles, especially in cooler gas-giant atmospheres. 

\subsection{Limitations of the model}

In this work we have opted to use a simple prescription of the atmosphere \citep{Guillot2010} that reproduces the structure of hot Jupiters without thermal inversions. 
Integral to the generation of the structure are the assumed mass absorption coefficients in the infrared and optical wavelength regimes (we assume $\kappa_\mathrm{ir} = 1 \times 10^{-2}$~cm$^2$~g$^{-1}$ and $\kappa_\mathrm{vis} = 4 \times 10^{-3}$~cm$^2$~g$^{-1}$, respectively). 
These determine the equilibrium pressure-temperature profiles, and it has been found that thermal inversions can appear when the optical opacity is much larger than that in the infrared \citep{Fortney2008, Guillot2010, Drummond2019}.  
Further, the atmospheric composition will influence the pressure-temperature profile \citep[][]{Molliere2015}, 
because the gas is typically the dominant source of opacity, as well as the dominant coolant.

In our simple model, we have ignored this coupling between the chemistry, the opacity, and the atmospheric structure.  
However, for the range of elemental abundances and C/O ratios that we have explored, significant differences in the 
pressure-temperature profiles are not expected, 
especially not in the deep atmosphere at around $0.1-1$~bar in a planet with a low effective temperature \citep[see, e.g., ][]{Molliere2015}. 
It is expected that the temperature here may increase by up to a few hundreds degrees only \citep[e.g.,][]{Madhusudhan2011, Drummond2019}.

On the other hand, if we were to consider an atmosphere with a thermal inversion due to the presence of strong absorbers in the optical, e.g., TiO and VO, the temperature in the upper atmosphere 
(<$10^{-2}$~bar), will be more affected which will influence the chemical composition therein and also the spectral features arising from this region 
\citep[e.g.,][]{Fortney2008}.
According to \citet{Madhusudhan2011}, TiO and VO will not cause a strong thermal inversion in a hot Jupiter atmosphere with C/O~$>1$, since TiO and VO are naturally underabundant for C/O~$ > 1$, so this will be more critical for oxygen-rich atmospheres which we predict to be more common.  

In addition, we have assumed the gas giant planet accretes gas directly from the disc midplane.  
According to recent theoretical \citep[e.g.,][]{Tanigawa2012, Morbidelli2014} and observational studies \citep{Teague2019}, the flow of gas into the gap which is opened in a disc by a growing planet is dominated by gas falling vertically from a height of at least one gas scale height. 
It is known that protoplanetary discs are not geometrically thin, and that they possess chemical layering due to heating and radiation from the central star \citep[e.g.,][]{Walsh2015}.
\citet{Cridland2020} compared the resulting atmospheric C/O ratios between planets accreting gas from the midplane and from between one and three scale heights after the gap has been opened by a giant planet. 
They concluded that when including such vertical accretion, the atmospheric C/O ratios tend to become lower, since more oxygen-rich icy dust grains become available for accretion onto the planetary atmosphere. 
Moreover, they find that the chemical composition of the gas dominates the final C/O ratios in planetary atmospheres if the planets are formed in the inner ($< 20$~au) region of the disc. 

Turning attention to our assumptions regarding the formation of the planet, we have assumed that all gas is accreted locally (at 0.5, 1, 5, and 20 au), and prior to the planet's radial migration to the final location of 0.05 or 0.1 au.
These assumptions are consistent with recent detailed planet accretion and migration models (e.g., \citealt{Mordasini2016, Cridland2019b}).
According to \citet{Mordasini2016}, \citet{Cridland2017, Cridland2019b}, for planets that have formed outside the water snowline, almost all of atmospheres will be captured just after core accretion is finished and before migrating to within the water snowline. The type II migration time scale ($>$a few $10^{6}$ year) is much larger than that of the runaway atmospheric acquisition \citep[$\ll 10^{6}$ year, e.g.,][]{Pollack1996}. 
Thus, planets accrete the bulk of their atmosphere from a narrow radial range during runaway acquisition of their atmospheres (e.g.,\citealt{Alessi2017, Cridland2019b}).

The disc midplane chemical models from \citet{Eistrup2016} that are used here considered a fixed set of global elemental abundances in gas \emph{and} ice in both the inheritance, and the reset scenarios, with C/O ratio of 0.34. This value was inspired by ISM ice abundances measured from infrared spectroscopy, but is below the solar value ($\sim$ 0.54, \citealt{Asplund2009}). Since the global C/O ratio of a protoplanetary disc should reflect that of its host star, it is possible that different protoplanetary discs around host stars with different elemental ratios will start out with different global C/O ratios. This might, in turn, lead to a different chemical evolution picture than what has been used here. Future work will explore the effects that changing the initial C/O ratio can have on the chemical evolution in the disc midplane, and further, which effects this may have on the evolution of the disc midplane C/O and O/H ratios (relevant for giant planet atmosphere formation) in gas and ice as function of radius.

Furthermore, in our calculations, we adopted the assumption of chemical equilibrium in the atmosphere justified by our focus on the denser regions ($\sim 0.1-1$~bar) where molecular emission is expected to originate.
The consideration of non-equilibrium chemistry (i.e., photo-processes, chemical kinetics, and vertical mixing) is expected not to change our main results; however, we acknowledge that photochemistry can enhance the HCN and \ce{C2H2} abundances in the atmosphere, and transport-induced quenching can enhance the abundances of \ce{CH4}, \ce{NH3}, and HCN in the lower atmosphere for wide range of C/O ratios \citep[e.g.,][]{Visscher2011, Moses2013, Madhusudhan2016, Tsai2017, Tsai2018, Hobbs2019, Molaverdikhani2019}.  
We intend to explore the impact of non-equllibrium chemistry in future work.

Finally in our model, we assume that the hot Jupiter has formed via the core accretion mechanism and that it has accreted its atmosphere locally and only from the gas in the vicinity of the planetary core \citep{Pollack1996, Ikoma2000}. However, there is another possible gas-giant planet formation mechanism that is disc fragmentation via the gravitational instability \citep{Durisen2007, Helled2014PPVI}.
In the latter mechanism, 
it is often assumed that the composition of the formed planet reflects that of the star (and bulk composition of the disc); however, recent work has shown that 
the relation between the formation location in the disc with respect to snowlines and elemental abundances of atmospheres can be more complicated than this assumption.
\citet{Ilee2017} calculated the physical and chemical structure of protoplanetary fragments in a gravitationally unstable disc, and found that molecular snowlines deviate significantly from the expected concentric ring structures found in axisymmetric discs. Increases in temperature caused by passing shocks desorb material at larger radii, and fragments that have formed develop surrounding snowlines.
In some fragments it is plausible for grains to sediment to the core before releasing their volatiles (e.g., \ce{H2O}) into the planetary envelope.
Thus, the atmospheric composition of planets formed via gravitational instability may not necessarily follow the bulk chemical composition (gas plus ice) of the disc from which they formed.

\section{Summary and conclusions}

In this study, we calculated the composition of hot gas-giant atmospheres assuming chemical equilibrium and using the open-source Thermochemical Equilibrium Abundances (TEA) code \citep{Blecic2016}.
We use elemental abundances (C/H, O/H, and N/H) extracted from chemical kinetics models of protoplanetary disc midplanes \citep{Eistrup2016}, in which different ionization rates and initial abundances had been explored. The aim was to investigate the relationship between chemical structure of gas-giant atmospheres and their formation conditions in chemically evolved protoplanetary discs, and to determine whether or not chemical evolution complicates the connection between the planet formation location relative to snowlines and the elemental composition of the atmosphere.

Similar to previous works, we find that as the value of the C/O ratio exceeds the solar value, the abundance of \ce{CH4} increases in the lower atmosphere ($\gtrsim 10^{-2}$~bar), and the abundances of \ce{C2H2} and \ce{HCN} increase mainly in the upper atmosphere ($\lesssim 10^{-1}$~bar). 
The abundances of oxygen-bearing molecules (e.g., \ce{H2O} and \ce{CO2}) correspondingly decrease. 
In contrast to previous work, we find that carbon-rich gas, i.e., C/O~$ > 1$ is achieved between the \ce{CO2} and \ce{CH4} snowlines in a protoplanetary disc that can exclude galactic cosmic rays and which has inherited interstellar gas and ice abundances only.  
In all other cases, the gas-phase C/O ratio remains $\lesssim 1$. 
This has two implications: first, if chemistry is active in disc midplanes then this can change the chemical fingerprint of the formation location of hot gas-giant planets over time and we expect that the accreted gas is oxygen rich, and second, the chemical imprint of snowlines in gas-giant atmospheres is only preserved for the case where chemistry is inactive in disc midplanes and in which interstellar abundances are wholly preserved in the protoplanetary disc. 
While the C/O ratio step function from \citet{Oberg2011} lays out radial regions of the disc midplane in which C/O ratios take distinct values in the gas and ice based on the locations of snowlines, we have demonstrated that using elemental abundances from chemical kinetics models of a disc midplane leads to wider diversity of atmospheric compositions than possible if only 
considering this simple picture.   
Nonetheless, we find that there exist unique combinations of O/H and C/O ratios at a chemical age of $\sim 1$~Myr that allow some constraints on the formation location of the planet, as well as providing insight into the chemical history of the material in the disc. 

Considering chemical evolution in the disc midplane over longer than Myr timescales can continue to alter the elemental C/O ratios of both gas and ice. 
This can further complicate connecting an exoplanet with constrained atmospheric composition to its formation history, including determining from where {\em and when} the planet has accreted disc material onto its atmosphere. 
Through examination of the evolution trends in the C/O ratios in the gas and ice we have shown in previous work that chemistry tends to process elemental carbon, oxygen and nitrogen from less volatile species, such as \ce{H2O}, \ce{CO2} and \ce{NH3} into more volatile species such as CO, \ce{O2} and \ce{N2} \citep[see, e.g.,][]{Eistrup2018}. 
This processing results in a disc midplane where carbon and oxygen-bearing molecules remain in the gas-phase out to the \ce{O2} snowline, where temperatures drop below $\sim 30$~K. 
This, in turn, means that an exoplanet accreting its atmosphere from gas inside the \ce{O2} snowline will simply accrete an atmosphere with the stellar C/O ratio (see Figure~\ref{Figure7_new2}). 
Potential solid icy impactors polluting this atmosphere may stem from outside the \ce{O2} snowline, and if these impactors come from outside the CO snowline (there will be C/O ratio variation in the ice between the \ce{O2} and the CO snowlines), then they will carry ices with stellar elemental abundances and the stellar C/O ratio, because all heavy elements beyond the CO snowline is in the form of ice. 
In summary, the fingerprint of snowlines are only retained in the atmospheres of planets assembled early ($\lesssim 1$~Myr) and from a disc in which negligible chemistry has occurred and preserve  
molecules which were formed in the molecular cloud phase. 
At the other scale, gas-giant planets that form within a chemically evolved disc midplane and at late times ($> $~ a few Myr), will give rise to an exoplanet atmosphere with stellar elemental abundances and the stellar C/O ratio.  Thus, reinforcing our prediction that carbon-rich gas-giant planets are likely to be rare.

Here we note that our conclusion that carbon-rich (C/O~$> 1$) hot Jupiters can only form between the \ce{CO2} and \ce{CH4} snowlines holds even if when considering the effects of efficient pebble drift. 
We acknowledge that the elemental abundances in the atmospheres (especially C/H) will be significantly increased relative to those predicted by chemistry-only models, which provides an observational diagnostic of the pebble drift phenomenon.  
However, in both cases, our conclusion that carbon-rich hot Jupiters can only form in a disc which has fully inherited interstellar abundances, and in which negligible chemistry has occurred, still holds.
In discs in which some degree of chemical reset has occurred (a more likely scenario), the formation of hot Jupiters with C/O~$>1$ (and also super-solar C/H ratios in many cases) 
is unlikely based on the calculations presented here and in \citet{Eistrup2016}.
However, we acknowledge that further modeling combining reset, chemistry, and pebble drift, are needed to test and quantify this.

For many hot Jupiters, the error bars on derived abundances from recent observations are too large (e.g., $>$0.5 dex in C/O ratios) to constrain characteristics related to their formation \citep{Line2014, Brewer2017}.
Using the next generation of facilities (such as JWST, ARIEL, SPICA, E-ELT, and TMT), it is anticipated that the determination of C/O ratios and elemental abundances will be possible for many hot Jupiter atmospheres, and with a precision (e.g., $<$0.2 dex in C/O ratios) that will allow distinguishment between the possible atmospheres proposed here (see also Figure~\ref{Figure7_new2}). 
According to \citet{Greene2016} and \citet{Schlawin2018}, using transmission and emission spectra observed by JWST with wavelengths longer than 2.5 $\mu$m, the C/O ratios can be constrained to better than 0.2 dex (corresponding to a factor of 1.6).
Here we note that \citet{Molliere2020} conducted atmospheric retrieval analyses for near-infrared spectra of directly imaged planet HR 8799e, part of which were recently obtained by VLT/GRAVITY \citep{GravityCollaboration2019}, and they constrained the C/O ratio of HR8799e's atmosphere with really impressive error bars: $0.68^{+0.07}_{-0.08}$.

We also predict that such observations will provide (some) insight into the degree of chemical processing in the disc prior to the onset of gas accretion when building gas-giant planets.
Recent ALMA observations have revealed multiple ring and gap structures in many protoplanetary discs \citep[e.g.,][]{Andrews2016, Andrews2018, Tsukagoshi2016, Huang2018, Isella2018, Notsu2019}, including in relatively young ($<$1 Myr) discs around Class I protostars, such as HL Tau \citep{ALMA2015} and AS 209 \citep{Fedele2018, Guzman2018}. 
Several theoretical studies have proposed that the planet-disc interaction causes material clearance within the orbits of newly-born gas-giant planets \citep[e.g.,][]{Goldreich1980, Kanagawa2015, Pinte2016}, providing hints that gas giants do form quickly, and sufficiently early in the protoplanetary disc lifetime for chemical fingerprints of snowlines to be retained in their atmospheres. 
\citet{Tanaka2019} discussed that pebble accretion would assist the early formation of gas giant planets in discs around Class 0/I protostars.
Important next steps will be to investigate, both observationally and theoretically, disc chemical structures at these earlier phases.

\section*{Acknowledgements}
We are grateful to Dr. John D. Ilee, Professor Tristan Guillot, Dr. Ingo Waldmann, Professor Giovanni Tinetti, Dr. Bun'ei Sato, and Mr. Daichi Fujita for useful comments on this work.
We thank the referee for many important suggestions and comments.
S.N.~is grateful for support from JSPS (Japan Society for the Promotion of Science) Overseas Research Fellowships, and RIKEN Special Postdoctoral Researcher Program (Fellowships). 
He is supported by MEXT/JSPS Grants-in-Aid for Scientific Research (KAKENHI) 16J06887 and 20K22376.
C.E.~acknowledges financial support from the Virginia Initiative on Cosmic Origins (VICO) Postdoctoral Fellowship program at the University of Virginia.
C.W.~acknowledges financial support from the University of Leeds and from the Science and Technology Facilities Council (grant numbers ST/R000549/1 and ST/T000287/1).
H.N.~is supported by MEXT/JSPS Grants-in-Aid for Scientific Research 18H05441, 19K03910 and 20H00182, NAOJ ALMA Scientific Research grant No. 2018-10B, and FY2019 Leadership Program at NAOJ.

\section*{Data availability}
The data underlying this article will be shared on reasonable request to the corresponding author.

\end{document}